\documentclass[]{aa}
\usepackage[varg]{txfonts}
\usepackage{graphicx}
\usepackage{amsmath}
\usepackage{amssymb}
\usepackage{lscape}
\usepackage{color}
\usepackage{longtable}
\usepackage{natbib,twoopt}
\usepackage{mathabx}
\usepackage{booktabs}
\usepackage{footmisc}

\newcommand{\OII}{[\mbox{O\,{\sc ii}}]}
\newcommand{\OIII}{[\mbox{O\,{\sc iii}}]}
\newcommand{\NeIII}{[\mbox{Ne\,{\sc iii}}]}

\newcommand{\NII}{[\mbox{N\,{\sc ii}}]}

\def\mean#1{\left< #1 \right>}

\begin{document} 

\title{Are LGRBs biased tracers of star formation? Clues from the host galaxies of the {\it Swift}/BAT6 complete sample of bright LGRBs. II: star formation rates and metallicities at $z < 1$\thanks{Based on observations at ESO, Program IDs: 077.D-0425, 177.A-0591, 080.D-0526, 081.A-0856, 082.D-0276, 083.D-0069, 084.A-0303, 084.A-0260, 086.A-0644, 086.B-0954, 089.A-0868, 090.A-0760, 095.D-0560}}

\author{
J. Japelj\inst{1,2}
\and
S. D. Vergani\inst{3,4,5} 
\and 
R. Salvaterra\inst{6}
\and 
P. D'Avanzo\inst{4}
\and 
F. Mannucci\inst{7}
\and
A. Fernandez-Soto\inst{8,9}
\and
S. Boissier\inst{10}
\and
L. K. Hunt\inst{7}
\and
H. Atek\inst{11,12}
\and
L. Rodr\'{i}guez-Mu\~{n}oz\inst{13,14}
\and
M. Scodeggio\inst{6}
\and
S. Cristiani\inst{1}
\and 
E. Le Floc'h\inst{15}
\and
H. Flores\inst{3}
\and
J. Gallego\inst{13}
\and
G. Ghirlanda\inst{4}
\and
A. Gomboc\inst{16,2}
\and
F. Hammer\inst{3}
\and
D. A. Perley\inst{17}
\and
A. Pescalli\inst{4}
\and
P. Petitjean\inst{5}
\and
M. Puech\inst{3}
\and
M. Rafelski\inst{18}
\and
G. Tagliaferri\inst{4}
}

\institute{INAF - Osservatorio Astronomico di Trieste, via G. B. Tiepolo 11, 34131 Trieste, Italy; \email{japelj@oats.inaf.it}
\and Faculty of Mathematics and Physics, University of Ljubljana, Jadranska ulica 19, SI-1000 Ljubljana, Slovenia
\and GEPI - Observatoire de Paris Meudon. 5 Place Jules Jannsen, F-92195, Meudon, France
\and INAF - Osservatorio Astronomico di Brera, via E. Bianchi 46, 23807 Merate, Italy
\and Institut d'Astrophysique de Paris, Université Paris 6-CNRS, UMR7095, 98bis Boulevard Arago, F-75014 Paris, France
\and INAF - IASF Milano, via E. Bassini 15, I-20133, Milano, Italy
\and INAF - Osservatorio Astrofisico di Arcetri, Largo E. Fermi 5, I-50125 Firenze, Italy
\and Instituto de Física de Cantabria (CSIC-UC), E-39005 Santander, Spain
\and Unidad Asociada Observatori Astronómic (IFCA - Universitat de Valéncia), Valencia, Spain
\and Aix Marseille Université, CNRS, LAM (Laboratoire d’Astrophysique de Marseille) UMR 7326, 13388, Marseille, France
\and Laboratoire d’Astrophysique, Ecole Polytechnique Fédérale de Lausanne,  Observatoire de Sauverny,  CH-1290 Versoix, Switzerland
\and Department of Astronomy, Yale University, 260 Whitney Avenue, New Haven, CT 06511, USA
\and Departamento de Astrofísica y Ciencias de la Atmósfera, Universidad Complutense de Madrid, Madrid E-28040, Spain
\and Dipartimento di Fisica e Astronomia “G. Galilei”, Università di Padova, Vicolo dell’Osservatorio 3, I-35122, Italy
\and Laboratoire AIM, IRFU/Service d'Astrophysique - CEA/DSM - CNRS - Université Paris Diderot, Bât. 709, CEA-Saclay, 91191, Gif-sur-Yvette Cedex, France
\and Faculty of Sciences, University of Nova Gorica, Vipavska cesta 11c, SI-5270 Ajdov\v s\v cina, Slovenia
\and Dark Cosmology Centre, Niels Bohr Institute, University of Copenhagen, Juliane Maries Vej 30, DK-2100 Copenhagen, Denmark
\and NASA Postdoctoral Program Fellow, Goddard Space Flight Center, Greenbelt, MD 20771, USA
}

\date{Received DD Mmmm YYYY / Accepted DD Mmmm YYYY} 
   
\abstract{}
 {Long gamma-ray bursts (LGRBs) are associated with the deaths of massive stars and could thus be a potentially powerful tool to trace cosmic star formation. However, especially at low redshifts ($z < 1.5$) LGRBs seem to prefer particular types of environment. Our aim is to study the host galaxies of a complete sample of bright LGRBs to investigate the impact of the environment on GRB formation.}
 {We study host galaxy spectra of the {\it Swift}/BAT6 complete sample of 14 $z < 1$ bright LGRBs. We use the detected nebular emission lines to measure the dust extinction, star formation rate (SFR) and nebular metallicity (Z) of the hosts and supplement the data set with previously measured stellar masses M$_{\star}$ (Paper I; Vergani et al. 2015). The distributions of the obtained properties and their interrelations (e.g. mass-metallicity and SFR-M$_{\star}$ relations) are compared to samples of field star-forming galaxies.}
{ We find that LGRB hosts at $z<1$ have on average lower SFRs than if they were direct star-formation tracers. By directly comparing metallicity distributions of LGRB hosts and star-forming galaxies, we find a good match between the two populations up to $12 + \log \left( \frac{\rm O}{\rm H}\right) \sim 8.4-8.5$, after which the paucity of metal-rich LGRB hosts becomes apparent. The LGRB host galaxies of our complete sample are not inconsistent with the mass-metallicity relation at similar mean redshift and stellar masses. The cutoff against high metallicities (and high masses) can explain the low SFR values of LGRB hosts. We find a hint of increased incidence of  starburst galaxies in the {\it Swift}/BAT6 $z < 1$ sample with respect to that of a field star-forming population. Given that the SFRs are low on average, the latter is ascribed to low stellar masses. Nevertheless the limits on the completeness and metallicity availability of current surveys, coupled with the limited number of LGRB host galaxies, prevent us from investigating more quantitatively if the starburst incidence is such as expected after taking into account the high metallicity aversion of LGRB host galaxies.
}
{}
 
 
\authorrunning{J. Japelj et al.} 
\titlerunning{Are LGRBs biased tracers of star formation? Clues from the BAT6 sample}
\maketitle
\section{Introduction}

Ever since long\footnote{GRBs are traditionally classified as long and short according to their observed duration (i.e. longer or shorter than $\sim 2$ s). In contrast to long GRBs, short GRBs are believed to arise from a merger of a compact object binary system and are found to have older progenitors \citep[e.g.][]{Fong2013}.} gamma-ray bursts (LGRBs) were first linked to the explosions of very massive stars \citep{Hjorth2003,Hjorth2012}, they have been considered as promising tracers of star formation in galaxies to very high redshifts \citep[e.g.][]{Kistler2008,Robertson2012,Perley2015a,Greiner2015}. LGRB host galaxies can be used as a complementary means to standard surveys of star-forming galaxies in order to understand galaxy properties and their evolution throughout cosmic history \citep[e.g.][]{Shapley2011,Carilli2013}. Studying LGRB hosts presents important observational advantages over studying luminosity selected galaxies. GRBs select galaxies independently of their brightness and thus avoid limitations (e.g. magnitude limited samples, dust extinction, redshift incompleteness) that usually accompany galaxy surveys. In particular, GRBs can pinpoint the faintest galaxies up to high redshifts ($z > 6$; \citealt{Tanvir2012,Basa2012,Salvaterra2013}), a population that might be a fundamental contributor to the re-ionization \citep{Salvaterra2011} but remains mainly elusive to conventional photometric and spectroscopic surveys.

In order to understand whether LGRB hosts can be used as a representative population of star-forming galaxies, we need to understand the link between the LGRB phenomena and star formation processes, in the following referred to as GRB (production) efficiency.  Of particular interest is the behaviour of GRB efficiency with respect to the properties of GRB host environment, such as stellar mass (M$_{\star}$), star formation rate (SFR) and metallicity. Studies in the past have reached contradictory conclusions regarding the LGRB efficiency, largely due to the heterogeneous nature of investigated samples \citep[e.g.][]{LeFloch2003,LeFloch2006,Fruchter2006,Savaglio2009,Levesque2010,Svensson2010,Mannucci2011,Graham2013,Perley2013,Hunt2014}. However, the large number of LGRBs detected by the {\it Swift} satellite \citep{Gehrels2004} accumulated in the past 10 years and carefully chosen selection criteria have recently resulted in several unbiased LGRB samples, highly complete in redshift: the GROND \citep{Greiner2011}, BAT6 \citep{Salvaterra2012}, TOUGH \citep{Hjorth2012b} and SHOALS \citep{Perley2015a} samples. With the help of these samples a more complete picture of the population of LGRB hosts is being revealed. 

At high redshifts the number of galaxies used in analyses is very small, which is presumably the reason why the conclusions drawn from different unbiased samples still differ: while some studies (\citealt{Greiner2015,Perley2015b}) claim that LGRB hosts can be direct tracers of star formation at about $z > 3$, others find the hosts to be of low luminosity with a metallicity-dependent efficiency \citep{Schulze2015}. 

The picture is gradually becoming more clear at low redshifts ($z < 1.5$). Several studies have investigated the metallicity of hosts and its effect on GRB efficiency, especially since theoretical models for single LGRB progenitor star have predicted low metallicity threshold above which LGRBs could not occur \citep{Yoon2006,Woosley2006}. The recent evidence, either direct \citep{Kruhler2015} or indirect \citep{Vergani2015,Perley2015b,Schulze2015}, from complete samples suggests that at low redshifts the LGRBs are indeed produced preferentially in low-metallicity environments. The metallicity threshold inferred from the data is $12 + \log \left( \frac{\rm O}{\rm H}\right) \sim 8.3-8.6$, confirming the findings of some of the previous studies focused on incomplete samples \citep{Modjaz2008,Levesque2010,Graham2013}. LGRB hosts at $z < 1$ are also found to be fainter and of lower stellar mass as compared to a field star-forming galaxy population (see also \citealt{Vergani2015,Perley2013,Perley2015b}). Because the stellar mass and metallicity of star-forming galaxies are correlated (stellar mass-metallicity relation, \citealt{tremonti2004}), the low-metallicity preference could provide the explanation for the differences in observed stellar masses between populations. Furthermore, owing to the fact that SFR and stellar mass of star-forming galaxies are correlated \citep[e.g.][]{Brinchmann2004}, metallicity has also been suggested as a possible explanation for the observed preference towards low SFRs in the LGRB host population (\citealt{Boissier2013,Kruhler2015}; but see \citealt{Michalowski2012}). However, metallicity may not be the only factor affecting the LGRB production efficiency \citep[e.g.,][]{Kelly2014,Perley2015c}. Even though a number of studies have addressed this issue, no self-consistent study has been performed simultaneously on stellar masses, SFRs and metallicities of a complete sample of LGRBs hosts. This is the goal of our study of the {\it Swift}/BAT6 complete sample of bright LGRBs.

Recently, \citet{Vergani2015} presented a study on the photometry and stellar masses of the $z < 1$ LGRB host galaxies of the {\it Swift}/BAT6 complete sample of bright LGRBs. The {\it Swift}/BAT6 sample \citep{Salvaterra2012} is selected according to favourable observing conditions \citep{Jakobsson2006} in order to avoid a biased selection. To ensure a significant redshift completeness -- the sample is 97 $\%$ complete in redshift -- LGRBs are furthermore selected by their brightness in gamma-rays ({\it Swift}/BAT peak flux $P \geq 2.6$ ph s$^{-1}$ cm$^{-2}$). The selection requirements do not depend on the brightness of optical afterglows, ensuring that the sample contains all LGRB population, including dark LGRBs \citep{Melandri2012}. \citet{Vergani2015} found that $z < 1$ LGRBs preferentially select faint, low-mass star-forming galaxies and are not unbiased tracers of star formation at $z < 1$. To better understand the interdependency of key properties of galaxies hosting LGRBs, here we expand the work of \citet{Vergani2015} by studying the emission line spectra of the hosts in the complete sample. Using the emission line fluxes, we measure the star formation rates and metallicities of the BAT6 sample hosts (Section \ref{analysis}). We compare the distributions of M$_{\star}$, SFR, and metallicity as well as their interrelations (i.e., SFR-M$_{\star}$, mass-metallicity MZ relation) to those derived from other samples of star-forming galaxies. Particularly, we focus our analysis on the completeness of the different comparison samples and the impact of different sample selection criteria on the final results (Section \ref{results}).

All errors are reported at 1$\sigma$ confidence unless stated otherwise. We use standard cosmology \citep{Planck2014}: $\Omega_{\rm m} = 0.315$, $\Omega_{\Lambda} = 0.685$, $H_{0} = 67.3$ km s$^{-1}$ Mpc$^{-1}$. All quantities are computed with respect to Chabrier initial mass function \citep{Chabrier2003}.

\section{Sample and data reduction}
\label{prepare}

Our sample is the same as presented in \citet{Vergani2015} and is composed of 14 $z < 1$ LGRBs of the {\it Swift}/BAT6 sample \citep{Salvaterra2012}. Because it is difficult to maintain a high level of GRB host data completeness at high redshifts \citep{Vergani2015}, we restrict ourselves to $z < 1$ range. In order to perform a study of the emission line properties of the host galaxies, we collected archival spectral observations of the hosts and carried out dedicated observational programmes to obtain the spectra of those hosts for which spectroscopic observations were missing. In the following subsections we detail our final spectroscopic data sets grouped by the instrument with which they were obtained. For the sake of homogeneity we reduced and analyzed the previously published data. In one case (GRB\,080319B) we detect neither continuum or emission lines. Our final sample therefore includes 13 host galaxies. The relevant information for each observation is summarized in Table \ref{tabobs}. 

\begin{table*}[ht]
\renewcommand{\arraystretch}{1.3}
\caption{Details of the observations of GRB hosts in the sample.}
\begin{center}
\scriptsize
\begin{tabular}{lccccccccccr}
\hline\hline
GRB host  & $E_{\rm B -V}^{G}$ & Date & Instrument & Exp. Time & Slit width$^{*}$  & Dispersion & Wavelength range & Seeing & Airmass & Reference\\
                & (mag)           &        &                   &      (s)            &      ($^{\second}$)      & ($\AA$/pixel) & ($\AA$)                 &     ($^{\second}$)       &              &  \\
\hline
050416A  & 0.026 & 2011-01-19 & VLT/X-shooter & 4$\times$900/4$\times$900/12$\times$300 & 1.0/0.9/0.9   & 0.4/0.4/1.0 & 3000--25000 & 0.7 & 1.5 & (1)\\
050525A  & 0.083 & 2012-09-18 & VLT/X-shooter & 4$\times$630/4$\times$664/4$\times$695     & 1.0/0.9/0.9JH & 0.4/0.4/1.0 & 3000--20000 & 1.5 & 1.6 & (1)\\
060614    & 0.019 & 2006-07-23 &  VLT/FORS2      & 1$\times$1500                                                   & 1.0                  & 3.3 & 3000--9600  & 1.2 & 1.2 & (2)\\
			    &           & 2007-04-26 & VLT/FORS1       & 2$\times$1200                                                   & 1.3                  & 1.5 & 3600--6000 & 0.8 & 1.5 & (3)\\
060912A & 0.046 & 2012-09-21 & VLT/X-shooter & 2$\times$900/2$\times$934/2$\times$965      & 1.3/1.2/1.2      & 0.4/0.4/1.0 & 3000--25000 & 0.7 & 1.5 & (1)\\
061021   & 0.051 & 2013-03-21 & VLT/X-shooter & 12$\times$940/12$\times$900/36$\times$320& 1.0/0.9/0.9JH   & 0.4/0.4/1.0& 3000--20000 & 0.9 &  1.1 & (1)\\
071112C$^{\dagger}$ & 0.105 & 2007-11-13 & VLT/FORS2       & 2$\times$1800                                     & 1.0             & 3.3 & 3000--9500 & 1.0 &  1.7 & (4)\\
080319B           & 0.010 & 2014-02-26 & GTC/OSIRIS       & 4$\times$825                                                     & 1.23            & 2.6 & 5100--10000 & 0.8 & 1.1 & (3)\\
                         &           & 2014-02-28 & GTC/OSIRIS      & 4$\times$825                                                     & 1.23             & 2.6 & 5100--10000 & 1.2 & 1.1 & (3)\\
080430             & 0.011 & 2014-01-07 & GTC/OSIRIS       & 8$\times$825                                                     & 1.23             & 2.6 & 5100--10000 & 1.1 & 1.1 & (3)\\
080916A           & 0.017 & 2008-09-17 & VLT/FORS1       & 6$\times$600                                                    & 1.0               & 3.3 & 3000--9500   & 0.8 & 1.2 & (4)\\
081007             & 0.014 & 2015-05-16 & VLT/X-shooter & 4$\times$700/4$\times$700/12$\times$250   & 1.3/1.2/1.2  & 0.4/0.4/1.0 & 3000--25000 & 1.0-2.1& 1.2& (3) \\
                         &           & 2015-06-20 & VLT/X-shooter & 8$\times$700/8$\times$700/24$\times$250   & 1.3/1.2/1.2   & 0.4/0.4/1.0 & 3000--25000 & 0.8-1.0& 1.1& (3)\\
                         &           & 2008-11-03 & VLT/FORS2       & 3$\times$2700                                                   & 1.0                & 3.2 & 5700--10000 & 0.5 & 1.1 & (5)\\
090424             & 0.022 & 2009-05-22 & VLT/FORS2       & 3$\times$3600                                                   & 1.0                & 3.2 & 6000--10200 & 0.7 & 1.4 & (5)\\
						  &           & 2013-04-08  & GTC/OSIRIS      & 3$\times$800                                                     & 1.23              & 2.1 & 3630--7500 &  0.9   & 1.1 & (5)\\
091018$^{\dagger}$   & 0.026 & 2009-10-18 & VLT/X-shooter  & 4$\times$600/4$\times$600/4$\times$600      & 1.0/0.9/0.9   & 0.4/0.4/1.0 & 3000--25000 & 0.9  & 2.0 & (6)\\
091127$^{\dagger}$   & 0.035 & 2010-02-12 & VLT/X-shooter  & 4$\times$1500/4$\times$1500/4$\times$750  & 1.0/0.9/0.9    & 0.4/0.4/1.0 & 3000--25000 &   1.0  & 1.1 & (7)\\
100621A           & 0.027 & 2012-10-16 & VLT/X-shooter  & 2$\times$1200/2$\times$1200/8$\times$300  & 1.0/0.9/0.9JH  & 0.4/0.4/1.0 & 3000--20000 & 0.9  & 1.3 & (1)\\
\hline
\end{tabular}
\tablefoot{For each GRB we report Galactic reddening in its line-of-sight and the information regarding the spectroscopic observation of its host galaxy: date of observation, telescope and instrument, instrumental setup (exposure time, slit widths, dispersion, wavelength range of the obtained spectrum), observing conditions (seeing and airmass) and the reference of the first published work presenting the spectrum. \newline $*$ 0.9JH stands for the X-shooter NIR slits with a special $K$-blocking filter \citep{Vernet2011}. \newline $\dagger$ Spectra are dominated by afterglow emission.\newline {\bf References}: (1) Kr\"uhler et al. 2015 (2) \citet{DellaValle2006} (3) This study (4) \citet{Fynbo2009} (5) \citet{Jin2013} (6) \citet{Wiersema2012} (7) \citet{Vergani2011}}
\label{tabobs}
\end{center}
\end{table*}

\subsection{VLT/X-shooter}

The X-shooter spectrograph \citep{Vernet2011} was used to observe eight hosts. For the purpose of this study we observed the GRB\,081007 host (Programme ID 095.D-0560, PI: S. D. Vergani). We also collected archival spectra of the hosts corresponding to GRBs\,050416A, 050525A, 061021 (PI: D. Malesani) 060912A, 091018, 091127 (PI: J. P. U. Fynbo) and 100621A (PI: T. Kr\"uhler). All observations were performed using the nodding technique with an offset of 5$^\second$ between individual exposures. Each observation included a telluric star, whose spectrum was taken right before or after the host's and at a similar airmass. A spectrum of a spectrophotometric standard star was taken at the beginning or end of the night. 

We processed the spectra using version 2.0 of the X-shooter data reduction pipeline \citep{Goldoni2006,Modigliani2010}. The raw frames were first bias subtracted and cosmic-ray hits were located and removed following the method of \citet{Dokkum2001}. The frames were divided by a master flat field. Day-time calibration frames were used to obtain a spatial-wavelength solution, necessary for the extraction and the rectification of orders. The rectified orders were shifted for the offset used in the observation and co-added to obtain a final two-dimensional spectrum, from which a one-dimensional spectrum with the corresponding error spectrum and bad-pixel map at the position of the source were extracted. In this way we reduced all observations, i.e. those of the host galaxies, telluric stars and spectrophotometric standards. Spectra of the latter were compared to tabulated flux calibrated spectra \citep{Vernet2010} to determine the response function which was then applied to the spectra of the hosts and telluric stars.

\subsection{VLT/FORS1 and FORS2}

From the ESO archive we collected the data of the hosts observed with the FORS1 and FORS2 instruments. To our knowledge the spectrum of the host of GRB\,060614 (FORS1; PI: J. Hjorth, Programme ID 177.A-0591(H)) has not been previously published. Already published data include hosts of GRBs\,060614 and 081007 (FORS2; PI: M. Della Valle), 071112C, 080916A (PI: P. Vreeswijk), and 090424 (PI: E. Pian).

The data were reduced using standard procedures for bias subtraction and flat-field correction.The extraction of the spectrum was performed with the ESO-MIDAS\footnote{\texttt{http://www.eso.org/projects/esomidas/}} software package. Wavelength and flux calibration of the spectra were achieved using a He-Ar lamp and observing spectrophotometric stars.

\subsection{GTC/OSIRIS}

We obtained GTC data with OSIRIS for the host galaxies of GRB\,080430 and
GRB\,080319B (programme GTC31-13B; PI: A. Fernandez-Soto). The former
were collected on January 7, 2014 and the latter over two nights on
February 25 and 27, 2014. In both cases the observing strategy was the same: a
brighter star was used as pivot and the target was centred on the slit by fixing the
OSIRIS rotation angle. A total of 6600 seconds (divided in 8 x 825 second
exposures) was integrated in each case, using a 15 arcsecond dithering motion
along the slit between each consecutive exposure. Conditions were good, with
clear dark sky, and seeing ranging between 0.8 and 1.2 arcseconds in different
exposures. In addition to GRB\,080430 and 080319B hosts, we collected observations 
of GRB\,090424 (PI: A. J. Castro-Tirado) from the archive.

The data were reduced using standard procedures and calibration files as
provided by the GTC. Wavelength calibration was obtained through the use of Hg-Ar,
Ne and Xe lamps that were observed during the same nights. A basic flux
calibration was obtained using the spectrum of the pivot stars and multi-band
photometry from SDSS.

\begin{table*}[!t]
\renewcommand{\arraystretch}{1.3}
\begin{center}
\small
\begin{tabular}{lccccccr}
\hline
\hline
GRB & $z$ & $\log {\rm M}_{\star}$&  $A_{\rm V}^{(a)}$ & \multicolumn{2}{c}{12 + $\log \left( \frac{\rm O}{\rm H}\right)$} & SFR & SF-tracer\\
\cmidrule(lr{.75em}){5-6}
       &        & M$_{\odot}$      &       (mag)              &                 M08 & KK04                          & (M$_{\odot}$yr$^{-1}$) & \\
\hline
050416A  & 0.6542 & 9.17$_{-0.12}^{+0.12}$ & 1.77$_{-0.56}^{+0.61}$ & 8.50$_{-0.15}^{+0.15}$ & 8.4$\pm0.2$ & 3.45 $\pm$ 1.42          & H$\alpha$\\
050525A$^{\dagger}$  & 0.6063 & 8.1$_{-0.6}^{+0.6}$       & -- &                --                    & --  & $>0.10$& \OIII \\
060614A  & 0.125 & 8.09$_{-0.17}^{+0.13}$ & 0.65$_{-0.34}^{+0.51}$ & 8.46$_{-0.20}^{+0.20}$   & 8.4$\pm0.2$ & 0.007$\pm$ -0.003        & H$\alpha$\\
060912A  & 0.9362 & 9.23$_{-0.07}^{+0.06}$ & 0.50$_{-0.25}^{+0.25}$ & 8.60$_{-0.12}^{+0.12}$ & 8.72$_{-0.08}^{+0.05}$ & 5.07 $\pm$ 0.93          & H$\alpha$\\
061021    & 0.3453 & 8.5$_{-0.5}^{+0.5}$       & 0.32$_{-0.32}^{+0.38}$ & 8.48$_{-0.26}^{+0.20}$ & 8.4$\pm0.2$ & 0.04 $\pm$ 0.01          & H$\alpha$\\
071112C  & 0.821 & 8.89$_{-0.18}^{+0.15}$ & $<$ 0.2                         & 7.90$_{-0.35}^{+0.50}$    & 7.88$_{-0.17}^{+0.27}$& 1.1 $\pm$ 0.4               & H$\gamma^{(b)}$\\
080430    & 0.767 & 8.15$_{-0.15}^{+0.12}$ & $<$ 0.1                          & 7.60$_{-0.35}^{+0.35}$   & 7.94$_{-0.21}^{+0.25}$& 1.65 $\pm$ 0.63           & H$\beta^{(b)}$\\
080916A$^{\dagger}$  & 0.688 & 8.98$_{-0.08}^{+0.07}$ & --  & 8.44$_{-0.22}^{+0.38}$  & 8.00$_{-0.20}^{+0.32}$ / 8.99$_{-0.41}^{+0.17}$& $> 0.20$ & \OII \\
081007    & 0.5294 & 8.78$_{-0.45}^{+0.47}$ & 0.49$_{-0.30}^{+0.37}$ & 8.32$_{-0.25}^{+0.20}$ & 8.32$_{-0.10}^{+0.10}$ / 8.58$_{-0.13}^{+0.10}$& 0.36 $\pm$ 0.07           & H$\alpha$\\
090424    & 0.5445 & 9.38$_{-0.19}^{+0.17}$ & 1.42$_{-0.51}^{+0.54}$ & 8.88$_{-0.20}^{+0.12}$ & 8.39$_{-0.13}^{+0.12}$ / 8.69$_{-0.13}^{+0.15}$ & 2.88 $\pm$ 1.14           & H$\alpha$\\
091018    & 0.9710 & 9.52$_{-0.10}^{+0.08}$ & 1.25$_{-0.75}^{+0.75}$ & 8.48$_{-0.30}^{+0.22}$ & 8.4 $\pm$ 0.2& 2.98 $\pm$ 1.81           & H$\alpha$\\
091127    & 0.4904 & 8.67$_{-0.07}^{+0.07}$ & $<$ 0.3                          & 8.50$_{-0.15}^{+0.15}$ & 8.47$_{-0.11}^{+0.12}$& 0.25 $\pm$ 0.02           & H$\alpha$\\
100621A  & 0.5426 & 9.04$_{-0.05}^{+0.06}$ & 0.34$_{-0.18}^{+0.21}$ & 8.40$_{-0.20}^{+0.15}$ & 8.25$_{-0.14}^{+0.16}$ / 8.67$_{-0.20}^{+0.14}$ & 8.92 $\pm$ 1.44           & H$\alpha$\\
\hline
\end{tabular}
\end{center}
\caption{Measured redshift, host extinctions, metallicites and star formation rates of our sample. Stellar masses are adopted from \citet{Vergani2015}. (a) Extinction was measured from Balmer decrement for all cases but GRBs\,050525A and 080916A - see text for details. (b) Flux of the significantly detected Balmer line was first transformed to the H$\alpha$ flux (see text and \citealt{Osterbrock2006}) and then to the SFR.\newline $\dagger$ Due to the lack of Balmer emission lines the host extinction cannot be computed. Reported star-formation rates are therefore formally lower limits.}
\label{tab_result}
\end{table*}

\subsection{Flux calibration verification}
\label{flux_cal}

Good flux calibration is essential to get reliable measurements of emission line fluxes. Flux calibrated host spectra were compared and cross-calibrated to photometric observations of the hosts \citep{Vergani2015}. In this way the slit losses were taken into account. However, there were a few exceptions. 

The host of GRB\,050525A has no detectable continuum and therefore we could not use the magnitudes to check the flux calibration. The X-shooter observations of telluric stars were obtained in similar conditions (airmass, seeing) and the same instrumental setup (binning, slit width) as the observations of scientific targets (e.g. hosts). We reduced the telluric star observation corresponding to the GRB\,050525A host using the same instrumental response function as for the science observations to flux-calibrate the telluric star spectrum. Then we calculated the flux correction by comparing the telluric star flux calibrated spectrum to photometric observations of the star. The same correction was applied to the host's spectrum. We note that we cross-checked this method for all other cases where both the host photometry and the telluric stars were available and the flux corrections obtained in this way were consistent within $\sim 20\%$ (see also \citealt{Piranomonte2015} and \citealt{Pita2014}). 

In the case of GRBs\, 071112C, 091018 and 091127 the spectra are dominated by afterglow emission. Flux calibration was therefore cross-checked by using photometric afterglow observations at (or near) the epoch in which the spectra were taken. We used light curves published by \citet{Wiersema2012} and \citet{Filgas2011} for GRBs 091018 and 091127, respectively. In the case of GRB\,071112C, the joint data sets of \citet{Huang2012} and \citet{Covino2013} were used.

\section{Analysis}
\label{analysis}

Emission line fluxes were measured by fitting one or multiple Gaussian functions to the data and cross-checked by integrating the signal under the line profile. Line fluxes (corrected for Galactic extinction, using extinction maps of \citealt{Schlafly2011}, and the average Milky Way extinction curve of \citealt{Cardelli1989}) are reported in Table \ref{line_flux}. Errors for each line were estimated by a Monte-Carlo simulation: for 1000 simulated events we repeatedly added random Gaussian noise (standard deviations were taken from the error spectra or rms of the continua) to the best-fit model and fitted the resulting spectrum by the same model. The obtained distribution of best-fit parameters was then used to compute the 1$\sigma$ errors. In case of a non-detection we calculated 3$\sigma$ upper limits by multiplying the rms in the region around the expected position of a line by 3. In cases where this resulted in particularly high upper-limits (e.g. GRB\,050525A host) we additionally checked the values by adding an artificial line to the spectrum -- assuming a Gaussian shape and {\it FWHM} as obtained from fitting strong lines of the same host -- and trying to measure it. In all these cases the artificial lines were not significantly detected, therefore we trust the upper-limits. 

Neither continuum or emission lines were detected in the case of GRB\,080319B host. The H$\alpha$ line was not covered by the GTC/Osiris spectrograph (see Table \ref{tabobs}), while strong \OIII\, and H$\beta$ fell in the region of strong telluric absorption. The host is faint ($r({\rm AB}) \sim 27$; \citealt{Tanvir2010}) therefore it is expected that the continuum was not detected. In the following we leave the host of GRB\,080319B out of the discussion, except when interpreting the impact that its absence has on the conclusions.

Balmer absorption lines are not clearly detected in our spectra, which is expected as LGRB hosts are faint young galaxies. The strength of the correction that should be applied to our measured line fluxes depends on several factors like stellar mass and spectral resolution \citep[e.g.,][]{Zahid2011}. Even though our sample spectra come with a wide range of spectral resolutions, the correction in all cases can be roughly approximated by the equivalent width of 1 $\AA$ \citep{Zahid2011,Cowie2008}, assuming the range of stellar masses of our sample \citep{Vergani2015}. The Balmer absorption correction is significant (i.e., larger than measured errors) only for the host of GRB\,090424. For others, while the correction has been added to the measured values, it is usually less than the uncertainty even if we assumed much larger equivalent line correction (e.g. 2 $\AA$).

\subsection{Extinctions, metallicities and star formation rates}

The measured rest-frame extinctions, star-formation rates (SFR) and metallicities are reported in Table \ref{tab_result}.

The host-integrated rest-frame extinctions $A_{\rm V}$ were determined from the Balmer decrement assuming gas with a temperature of $T = 10^{4}$ K (i.e. intrinsic ratios between different hydrogen Balmer lines are assumed to be H$\alpha$/H$\beta$ = 2.87, H$\gamma$/H$\beta$ = 0.47 and H$\delta$/H$\beta$ = 0.26; \citealt{Osterbrock2006}). To measure the extinctions we used only lines detected with 3$\sigma$ confidence and assumed the Milky Way\footnote{We chose the Milky Way extinction curve because it is commonly used in the literature. We note that applying other commonly used extinction curves \citep[e.g.][]{Japelj2015} does not result in a significant difference in the measurement of $A_{\rm V}$ and subsequent correction of line fluxes.} extinction curve \citep{Pei1992}. However, the hosts of GRB\,050525A and GRB\,080916A lack the Balmer lines needed to measure the extinction. While the line-of-sight extinction, measured from the afterglow spectral energy distribution, is available for the two cases, in general line-of-sight and host-integrated extinctions are not necessarily the same \citep[e.g., see Section \ref{secdust} and][]{Perley2013}. Therefore we assumed $A_{\rm V} = 0$ in the case of these two hosts in the further analysis.

All the steps described in the following paragraphs were performed after applying the host extinction correction to the emission lines.

To measure the SFR, where possible, we used the H$\alpha$ line because it is the most reliable tracer of SFR and it does not depend strongly on the uncertainties in the measured extinction. We assumed the conversion between H$\alpha$ luminosity and SFR as given by \citet{Kennicutt1998}, but scaled to the \citet{Chabrier2003} initial mass function. In two cases (hosts of GRBs\,071112C and 080430) we scaled other significantly detected (and extinction corrected) Balmer lines to H$\alpha$ (assuming intrinsic ratios between Balmer lines) and used the same prescription to derive the SFRs. None of the Balmer lines is significantly detected in the case of GRB\,080916A and 050525A hosts, therefore we used the \OII\, and \OIII$\lambda5007$ lines as SFR tracers for the two hosts, respectively. \OII\, luminosity is known to be strongly correlated with SFR in the LGRB host samples \citep{Savaglio2009,Kruhler2015}. \citet{Kruhler2015} also found a correlation between L(\OIII$\lambda5007$) and SFR, although the relation is quite scattered (the scatter of the relations was taken into account in the final estimation of the errors). We cross-checked the SFR-L(\OII) and SFR-L(\OIII$\lambda5007$) relations found by \citet{Kruhler2015} on our sample, using 9 GRB hosts with simultaneously detected H$\alpha$, \OII\, and \OIII$\lambda5007$ lines. We found nearly the same relations (and therefore almost identical calculated SFRs for the GRB\,080916A and 050525A hosts) but with a bit larger scatter. We also verified that the marginally detected H$\alpha$ line in the case of GRB\,050525A gives a nearly identical value of SFR as \OIII$\lambda5007$. Because host extinction for the hosts of GRB\,080916A and GRB\,050525A is unknown, the measured SFRs are formally lower limits.

Gas phase metallicities of distant galaxies are typically measured using strong emission line ratios, whose dependence on metallicity has been determined either via theoretical models or cross-calibration with direct metallicity measurements in the local Universe (e.g. see review by \citealt{Kewley2008}). We decided to measure metallicities by using the method of \citet{Maiolino2008} (see also \citealt{Mannucci2011}), where gas phase metallicities were computed by simultaneously minimising all metallicity indicators that can be used for each specific case. In principle, the method has two free parameters: host extinction and metallicity. However, since most of the indicators are built from ratios of lines of similar wavelengths, they are not sensitive to extinction and therefore that parameter is largely unconstrained in the minimisation procedure. We therefore fixed the extinction values as obtained from the Balmer line ratios. We determined the metallicities for all cases but GRB\,050525A. Even though we lack the host extinction measurement for the host of GRB\,080916A, we do not expect that to have a significant effect on the metallicity measurement and the final conclusions (given the high metallicity errors measured in this case), unless extinction turned out to be very high ($A_{\rm V} > 3$ mag). Such high value of host average-extinction at $z < 1$ is very unlikely (see Figure 11 in \citealt{Perley2013}). In order to prove that our conclusions do not depend on the choice of the assumed indicator, we also determined metallicities\footnote{We used also the pyMCZ software \citep{Bianco2015}} using the $R_{\rm 23}$-based calibration of \citet{Kobulnicky2004} (KK04). The calibration suffers from degeneracy, that is, for each measured line ratio one obtains two metallicity solutions. The degeneracy can be broken with the help of metallicity-dependent \NII/\OII\, or \NII/H$\alpha$ ratios. Several cases of our hosts have the KK04 metallicities near the turnover point at $12 + \log \left( \frac{\rm O}{\rm H}\right)_{\rm KK04} \sim 8.4$ and for these we assumed the metallicity of 8.4 and added an error of 0.2 dex. The metallicity solution for the hosts of GRBs 071112C and 080430 was also double valued with two extreme lower- and upper-branch values (12 + $\log \left( \frac{\rm O}{\rm H}\right)=7.88$, 8.91 and 7.94, 8.87, respectively). For both hosts we detect the \NeIII\, emission line. The \NeIII/\OII\, diagnostic \citep{Maiolino2008}, definitely points to a low metallicity for both hosts (12 + $\log \left( \frac{\rm O}{\rm H}\right)=7.90$ and 7.50, respectively). We therefore choose the lower branch solution for these two galaxies. We note that the reliability of diagnostics, which are based on \OIII\, and \NII\, emission lines, at high redshifts have been questioned \citep{Kewley2013,Shapley2015}. However this uncertainty should not affect our $z < 1$ sample study. 

\section{Comparison star-forming galaxy samples}
We want to compare the SFR and metallicity properties of our LGRB host galaxy sample with sample(s) of a field population of star-forming galaxies. In order to make the comparison reliable, spectroscopic surveys are necessary. In addition, the completeness limits of the surveys (in terms of brightness, SFRs and stellar masses) should be deep enough for a valid comparison with our BAT6 sample. It is difficult to find surveys with such characteristics. The best survey turned out be the VIMOS VLT Deep Survey \citep[VVDS;][]{LeFevre2005}, especially because its magnitude selection is deep enough to cover the faint magnitudes of the hosts in our sample. We adopt the VVDS sample as the primary comparison sample. As explained in detail in the following, a few other surveys were also used for different tests.

In the followings, all the stellar masses were scaled to the \cite{Chabrier2003} IMF.

\subsection{VIMOS VLT Deep Survey}
\label{vvds}
The VVDS is a comprehensive survey of $z < 6.7$ star-forming galaxies conducted with the VLT/VIMOS multi-object spectrograph \citep{LeFevre2003}. We retrieved the last data release \citep{LeFevre2013} from the VVDS-database\footnote{http://cesam.lam.fr/vvdspub/}. In particular, we selected the data corresponding to $0.1 < z < 1.0$ star-forming galaxies collected in magnitude limited {\it Deep} ($17.5 \leq i_{\rm AB} \leq 24$) and {\it Ultra-Deep} ($23 \leq i_{\rm AB} \leq 24.75$) surveys. The latter covers an area on the sky included in the former field. The combined sample consists of a total of 6366 galaxies with measured stellar masses and host extinction. From this sample, we selected galaxies with detected emission lines. In particular, for the purpose of calculating the emission-line-based SFR, we required a detection of at least one of the following lines: \OII, H$\alpha$ or H$\beta$ with significance > 2$\sigma$. This requirement reduced the number of galaxies in VVDS sample to 3551. The properties of this sample (i.e. distributions of apparent $i_{\rm AB}$ and absolute $M_{\rm B}$ magnitudes, stellar masses and redshift) do not differ significantly from the original sample, therefore, we did not introduce any additional bias with these selection requirements (see Figure \ref{plot11} in the Appendix).

\begin{figure}[]
\centering
\includegraphics[scale=0.48]{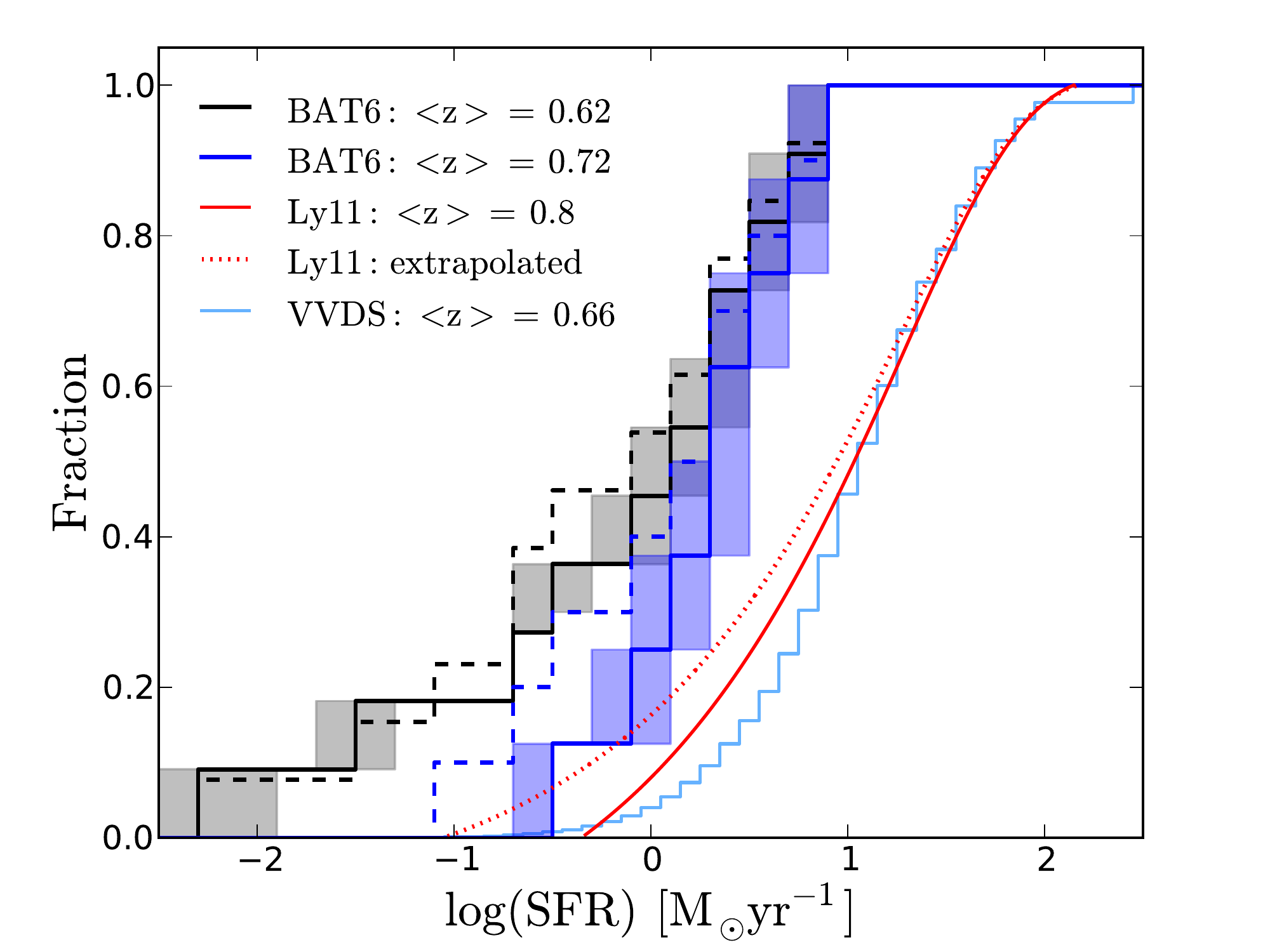}
\caption{Cumulative SFR distributions of our sample (solid black line) and its $z > 0.5$ subsample (solid blue line). Shaded regions show 1$\sigma$ sampling range around solid lines. Dashed lines show distributions including lower limits. For comparison we also plot a star-formation weighted distribution of $z = 0.8$ star-forming galaxies \citep{Ly2011} (red solid line), the same distribution extrapolated towards lower SFRs to account for completeness limit of the survey (red dotted line), and the $z >0.3$ VVDS sample (light blue line; see text for details).}
 \label{fig1}
\end{figure}

Star formation rates were calculated in the following way. Line fluxes were corrected for host extinction. Following our procedure for BAT6 hosts, we calculated SFRs from the H$\alpha$ or H$\beta$ line. If neither of the two was available (or was detected with a very high uncertainty), \OII\, was used to measure the SFR. In the latter case, we used the calibration between SFR, \OII\, luminosity and intrinsic brightness $M_{\rm B}$ of the hosts given by \citet{Moustakas2006}. Once the SFRs were obtained, we compared the sample properties to some of the other samples in order to cross-check whether our selection is unbiased and to better understand the completeness limits\footnote{When comparing cumulative distributions we compare LGRB host properties (e.g. SFR or metallicity) with SFR-weighted properties of field star-forming galaxies. We therefore assume that the probability of hosting a LGRB is proportional to the SFR of a galaxy. If the properties of the two populations differ, then the initial assumption that we are testing is wrong and we try to understand which are the factors that make it wrong.}. We found that the SFR-weighted mass distribution of the SFR-selected VVDS sample agrees very well with the UltraVista sample \citep{Ilbert2013} used by \citet{Vergani2015} as a reference sample of masses of field galaxies. Second, the SFR-weighted SFR distribution agrees quite well with the SFR-weighted distribution built from the H$\alpha$ luminosity function of \citet{Ly2011} (hereafter Ly11, see Figure \ref{fig1} and section 3.2.2). The SFR completeness limit of the VVDS and Ly11 surveys is similar with $\log {\rm SFR} ~{\rm [M}_{\odot} {\rm yr}^{-1}{\rm ]} \sim 0.0$.

We determined metallicities using two different methods. First, we used the same approach as for the BAT6 hosts, i.e. simultaneously minimizing a number of different line ratios corresponding to different calibrators. Unfortunately only a portion of the VVDS sample has enough emission lines detected to provide a reliable metallicity determination (if other indicators are used this portion is even smaller). Upon examination we found that the subsample for which we could measure the metallicity was slightly biased towards low stellar masses and therefore low metallicities. 
As the VVDS galaxies represent a field population of star-forming galaxies, they should follow the fundamental mass metallicity relation \citep[FMR;][]{Mannucci2010,Mannucci2011}. Using the stellar masses and the previously measured SFRs for the SFR-selected VVDS sample one can therefore calculate the metallicities from the FMR\footnote{In the calculation of the FMR-based metallicity we account for errors in measured SFR and stellar mass as well as intrinsic dispersion of the FMR relation ($\sim 0.06$ dex). The effect of the latter is negligible.}. As expected, the metallicites calculated from the two methods do not differ statistically for the aforementioned subsample. In the following we therefore use the FMR-based metallicities.

\subsection{NEWFIRM H$\alpha$ survey}
The majority of the SFRs determined for the VVDS sample is based on the luminosity of the \OII\, line, for which the strength of the line is sensitive to the abundance and ionization state of the gas \citep{Kewley2004}, making the \OII\,-SFR relation rather controversial (e.g. \citealt{Moustakas2006} and references within), especially for heterogeneous samples of galaxies.  We therefore additionally used the NEWFIRM H$\alpha$ survey of star-forming galaxies Ly11 for the purpose of comparing the SFR distributions of LGRB hosts and star-forming population.

The Ly11 field galaxy SFR distribution is the results of the NEWFIRM narrowband H$\alpha$ observational campaign. By observing a sample of $\sim 400$ star-forming galaxies at $z \sim 0.8$, Ly11 built an H$\alpha$ luminosity function at this redshift. 
For the comparison with the LGRB host galaxies, we multiplied the Ly11 luminosity function (described by a Schechter function with $\log (L_{*}/({\rm erg} ~{\rm s}^{-1})) = 43.00$ and $\alpha = -1.6$) by the luminosity to account for the assumption that the probability of hosting a LGRB is proportional to the SFR of a galaxy. Luminosities were then converted to SFRs (following \citealt{Kennicutt1998}, but scaled to \citealt{Chabrier2003} IMF).

\subsection{Other star-forming galaxy samples}
\label{othersample}

Most of the LGRB host galaxies in our $z<1$ sample have stellar masses below 10$^9$\,M$_{\sun}$, and quite high specific SFR (see Section \ref{sfrs}). We therefore considered for comparison also samples focused on those kind of star-forming galaxies to see if they have similar properties than LGRB host galaxies. We note that the following two samples are not unbiased, as they were both selected to address specific star-forming populations.

\cite{Atek2014} presented the properties (stellar masses and SFRs) of 1034 galaxies at $0.3<z<2.3$ selected through emission lines with the WISP \citep{Atek2010} and 3DHST \citep{Brammer2012} surveys. This selection favours the detection of starburst galaxies. We use the $0.3<z<1$ sub-sample properties for comparison with those of the LGRB host galaxies in our sample (see Section \ref{sfrs}).

We also use the sample of 45 low-mass star forming galaxies and 29 blue compact dwarf galaxies (BCD, defined following \citealt{Gildepaz2003}) studied by \cite{Rodriguez2015}, to check if the GRB host galaxies have in some way similar properties to those classes of galaxies. These samples have been selected at first using photometry but then a spectroscopic redshift determination is required, hence implying the detection of emission lines.

\section{Results and Discussion}
\label{results}

\subsection{Star formation rates and stellar masses}

\subsubsection{Star formation rates}
\label{sfrs}

The cumulative SFR distribution of our 13 LGRB hosts (Table \ref{tab_result}) is shown in Figure \ref{fig1}. Due to the uncertainty in the estimated host extinction the SFR errors are quite high in some cases. For this reason we also plot the 1$\sigma$ uncertainty region (shaded), obtained by performing MC simulation in which the distributions are generated from the SFRs, varied by the measured error. In the same plot we show the SFR-weighted cumulative distributions of the VVDS and Ly11 samples. As illustrated in Figure \ref{fig1}, LGRB hosts can hardly be drawn from the star-formation weighted distribution of star-forming galaxies. The average redshifts of BAT6, VVDS and Ly11 samples are $\mean{z} = 0.62, 0.7$ and 0.8, respectively. The SFR density of field galaxies is observed to evolve with redshift \citep[e.g.][]{Whitaker2012,Speagle2014}. In order to remove a systematic difference due to the effect of the observed evolution, we cut the BAT6 sample including only $ 0.5 < z < 1$ hosts (excluding lower limits, this leaves us with a sample of  8 hosts) with $\mean{z} = 0.72$.  It is evident from Figure \ref{fig1} that in this way we cut the low end of the SFR distribution.

\begin{figure*}[t]
\centering
\begin{tabular}{cc}
\includegraphics[scale=0.45]{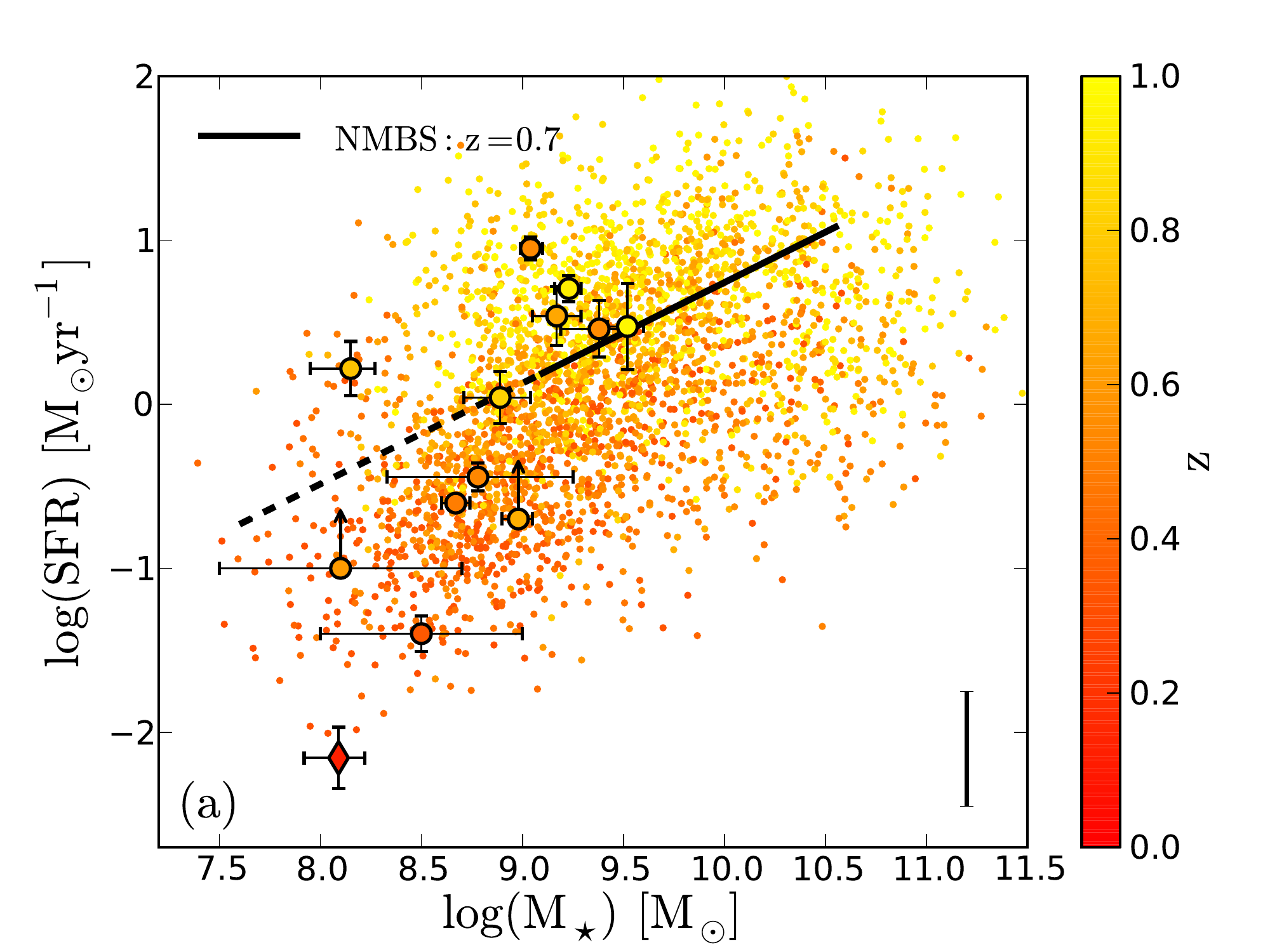}&
\includegraphics[scale=0.45]{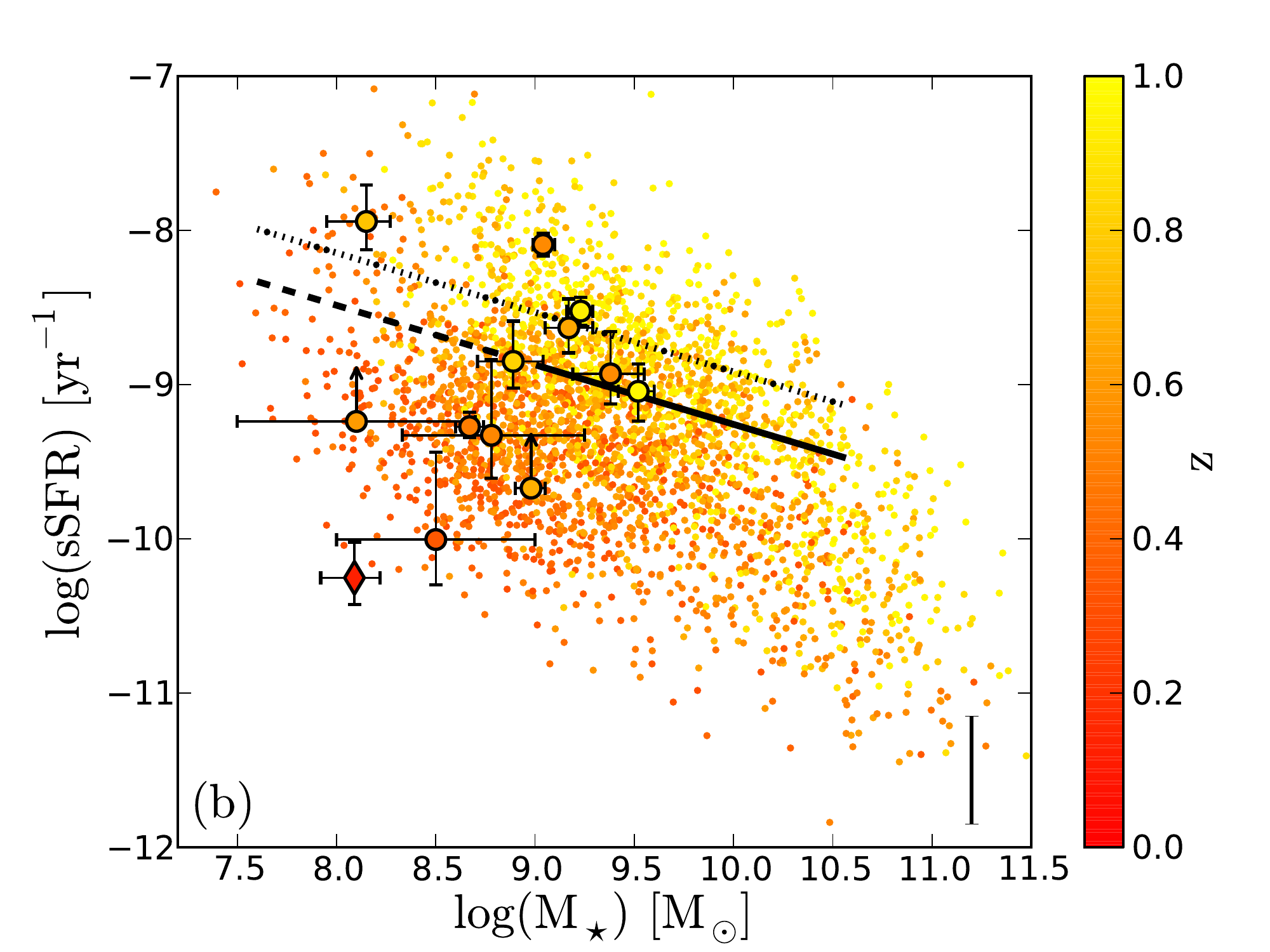}\\
 \end{tabular}
\caption{{\it (a)} SFR-stellar mass relation for BAT6 sample. The host of the GRB\,060614A is plotted with a different symbol (diamond) to emphasise the dubious nature of the GRB. The color-coding corresponds to redshifts as noted with the color bar on the right sight of the plot. Small points with the same color-coding correspond to $0.3 < z < 1.0$ VVDS survey of star-forming galaxies \citep{LeFevre2013}. In addition, we plot the median value of SFR-stellar mass relation at $z \sim 0.7$ (mean redshift of the VVDS sample and the BAT6 sample without the host of GRB\,060614) as observed in the NEWFIRM Medium Band Survey \citep[NMBS;][]{Whitaker2012}.  Note that the the latter relation has a scatter of $\pm 0.34$ dex (indicated by an errorbar in the plots). With a dashed line we draw the extrapolation of the relation below the stellar mass completeness of the \citet{Whitaker2012} survey.
{\it (b)} Specific SFR-mass relation. 
The median value of the \citet{Whitaker2012} relation at $z \sim 0.7$ is plotted. The dotted line represents the relation plus the dispersion (0.34 dex).}
 \label{fig2}
\end{figure*} 

\begin{figure*}[t]
\centering
\begin{tabular}{cc}
\includegraphics[scale=0.45]{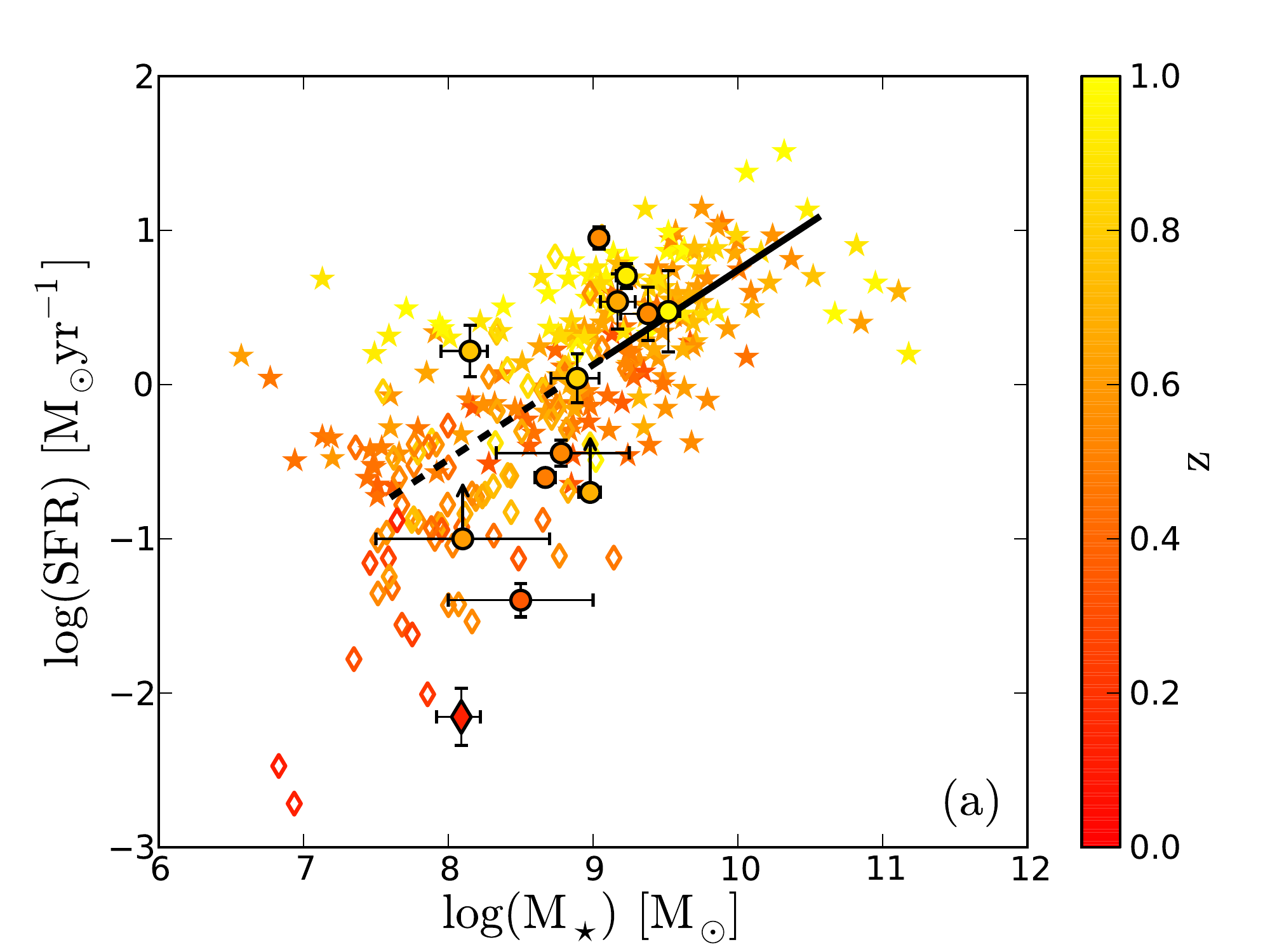}&
\includegraphics[scale=0.45]{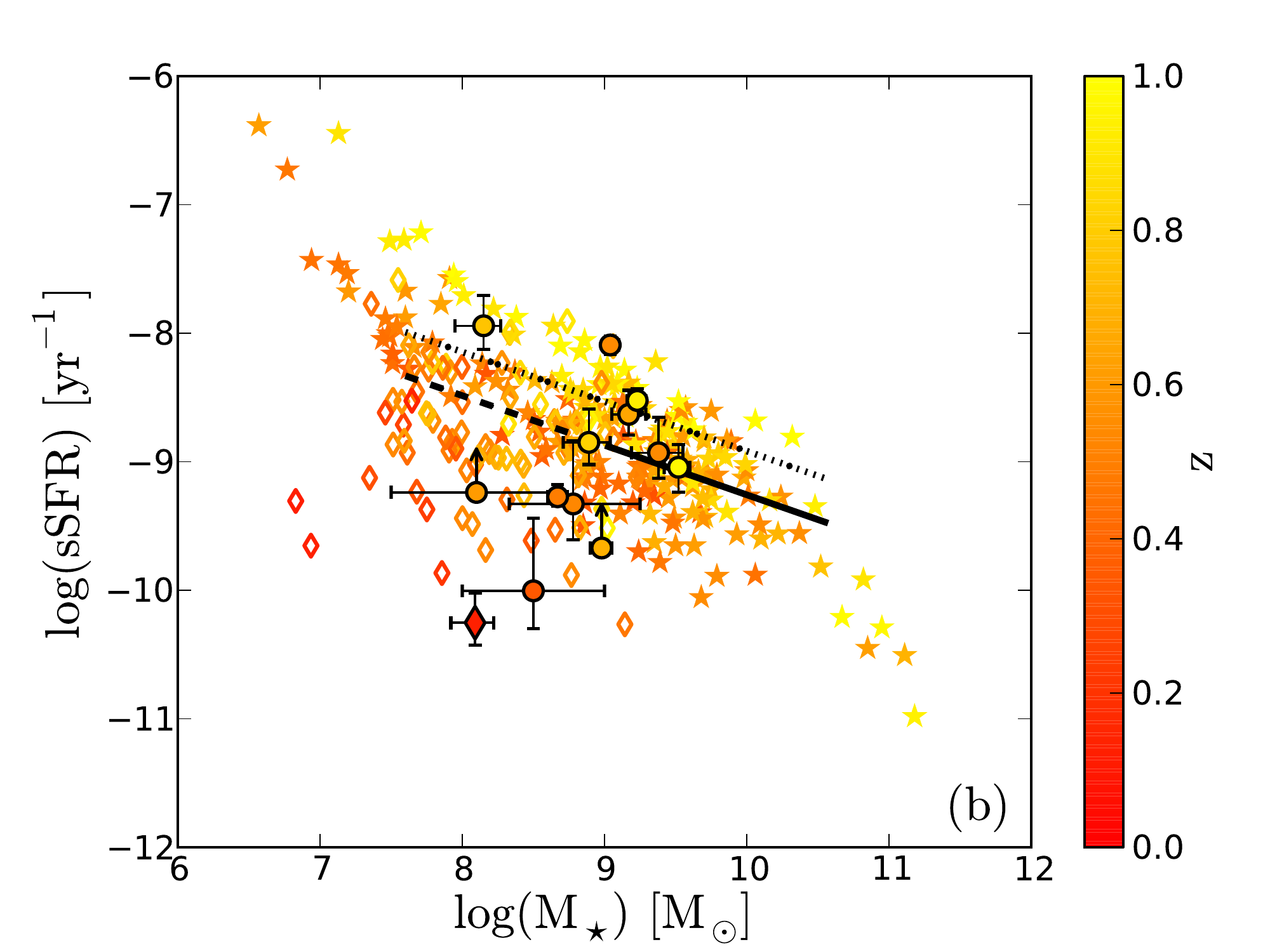}\\
 \end{tabular}
\caption{Comparison of {\it (a)} SFR-stellar mass and {\it (b)} sSFR-stellar mass relations of our BAT6 sample to the samples of extreme starbursts \citep[star symbols;][]{Atek2014} and blue compact dwarf galaxies \citep[non-filled diamonds;][]{Rodriguez2015}. The color scale and the overplotted lines are the same as in Figure \ref{fig2}. The host galaxies of GRB\,060614A and 061021, while included in the plots, were excluded from the comparison of specific SFRs (see text) because their measured SFR is below the completenes limit of the two surveys ($\log ~{\rm SFR} {\rm [M}_{\odot} {\rm yr}^{-1}{\rm ]}\sim-1$). }
 \label{fig3}
\end{figure*} 

To quantify the difference between the two samples we performed a Kolmogorov-Smirnov (KS) test, testing whether the BAT6 (cut sample, where lower limits are not taken into account) and Ly11 samples\footnote{It does not affect the conclusions whether we make a comparison with VVDS or Ly11 samples.} are drawn from the same distribution. We ran a MC simulation, in which we randomly chose SFRs from BAT6 sample (varying the measured values by their errors) and a number of 400 SFR values (number of galaxies in Ly11 sample), randomly chosen from the Ly11 distribution (Figure \ref{fig1}). This resulted in a probability of $p \approx 0.007$, which suggests that we can discard the hypothesis. This test has been performed considering the Ly11 luminosity function only in the range in which the completeness limit of the Ly11 sample is trustworthy (i.e., down to H$\alpha$ $ = 10^{41}$ erg cm$^{-2}$ s$^{-1}$, which corresponds to $\log ~{\rm SFR} {\rm [M}_{\odot} {\rm yr}^{-1}{\rm ]}\sim 0$). However, as seen in Figure \ref{fig1}, the SFRs of the BAT6 sample extend to lower values. To account for this discrepancy, we make two additional tests. First we cut the BAT6 sample to the same SFR completeness limit, resulting in the probability of $p \approx 0.015$. Alternatively, we assume that the H$\alpha$ luminosity function can be simply extrapolated to lower luminosities\footnote{The cumulative distribution depends on the assumed SFR limit, i.e. it will change if we extrapolate the distribution down to lower SFRs. Our conclusion is therefore sensitive to the assumed limit. However, even if we extrapolate the luminosity function down to $\log {\rm SFR} {\rm [M}_{\odot} {\rm yr}^{-1}{\rm ]} = -2$, the cumulative distribution does not change a lot and the significance of the discrepancy remains similar. This is because the distribution is weighted for SFR.} (down to H$\alpha$ $ = 10^{40.3}$ erg cm$^{-2}$ s$^{-1}$ to match the lowest value of our cut BAT6 sample; dotted line in Figure \ref{fig1}). Using the extrapolated distribution, the KS test gives the values of $p \approx 0.02$. We can therefore conclude that LGRB formation is more efficient in low SFR environment (see also the similar comparison and results found by \citealt{Kruhler2015} for their X-shooter sample of GRB host galaxies).

\subsubsection{SFR vs stellar mass relation}
\label{sfms}

A correlation between the SFR and the stellar mass, known as the 'star formation main sequence' (SFMS), has been found to exist for star-forming galaxies in the full range from low \citep[$z < 1$;][]{Brinchmann2004} to high  \citep[$z \sim 6$;][]{Steinhardt2014} redshifts. Both the slope and normalisation of the correlation are observed to change over cosmic time \citep[e.g.][]{Speagle2014}. In order to asses whether GRB hosts occupy the same  SFR-M$_{\star}$ region as the field star-forming galaxy population, we plot our BAT6 sample in the SFR-M$_{\star}$ plane (Figure \ref{fig2}a). We compare our values to the star-forming galaxies from the VVDS survey. 

In general, SFR of the BAT6 sample is increasing with stellar mass as expected. In agreement with the results of PaperI, there is a clear discrepancy on the stellar mass range covered by the VVDS and the LGRB host galaxies, the first extending to much higher stellar masses. Within the LGRB stellar mass range, while the values for GRB hosts are quite scattered they occupy the same region as VVDS field galaxies (at similar redshifts). Two low-redshift hosts (corresponding to GRBs\,060614A and 061021) stand out with very low values of both SFR and specific SFR. We caution, however, that GRB\,060614A is rather peculiar in itself, because even though its duration clearly makes it a long GRB, no supernova (SN) has been detected at the position of the burst, despite its near origin and a comprehensive follow-up campaign \citep{Fynbo2006, DellaValle2006}. In fact, recently re-analysed late-time data of this GRB afterglow show evidence \citep{Yang2015,Jin2015} of an emergent macronova emission \citep{Li1998}, the radioactive decay of debris following a compact binary merger. Thus the origin of this GRB is most likely different from the other LGRBs in the sample. This does not affect our results, because this GRB host is excluded from every comparison in the following as it never satisfies the completeness limits of the surveys.

\begin{figure*}[t]
\centering
\begin{tabular}{cc}
\includegraphics[scale=0.45]{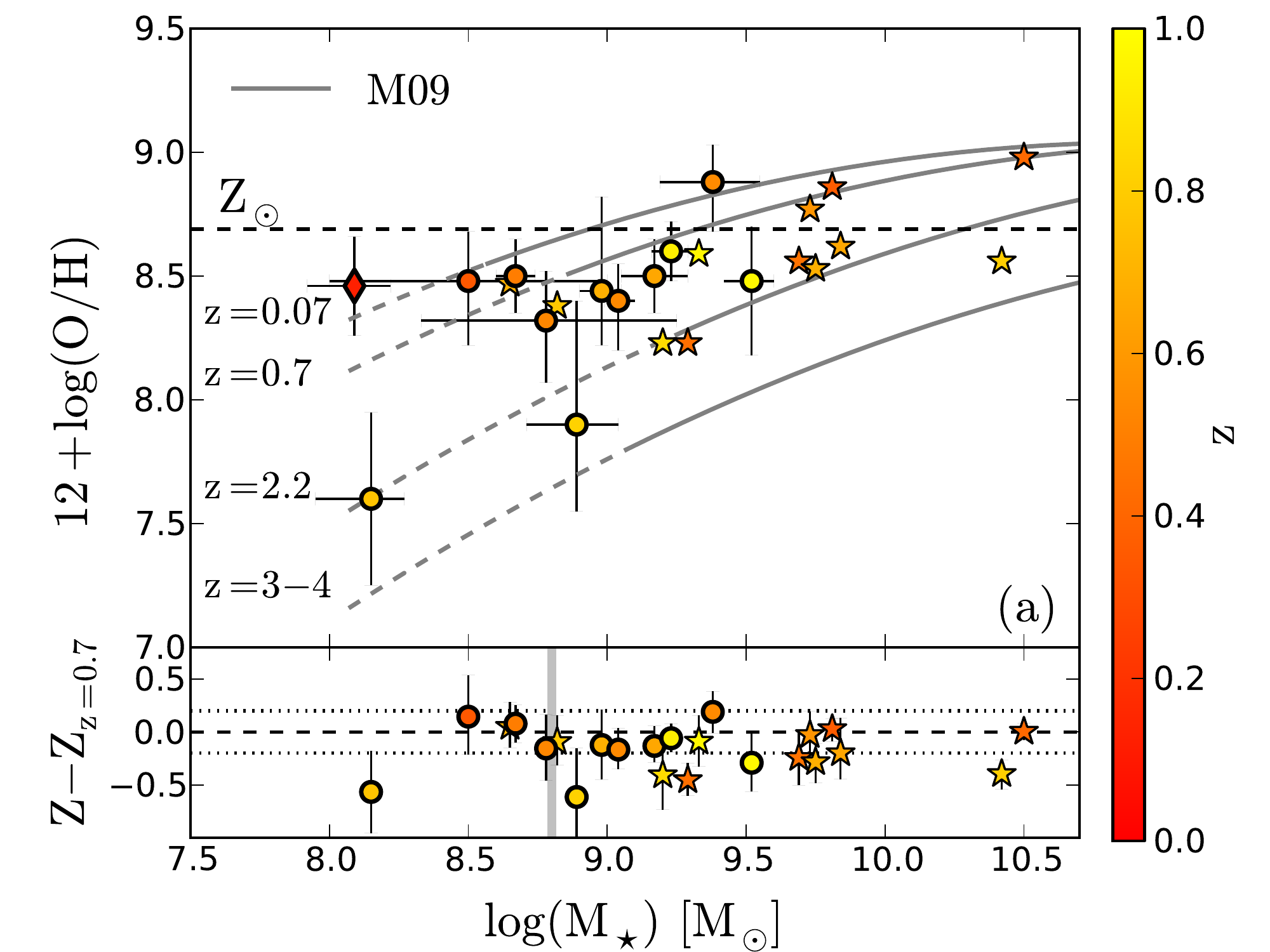}&
\includegraphics[scale=0.45]{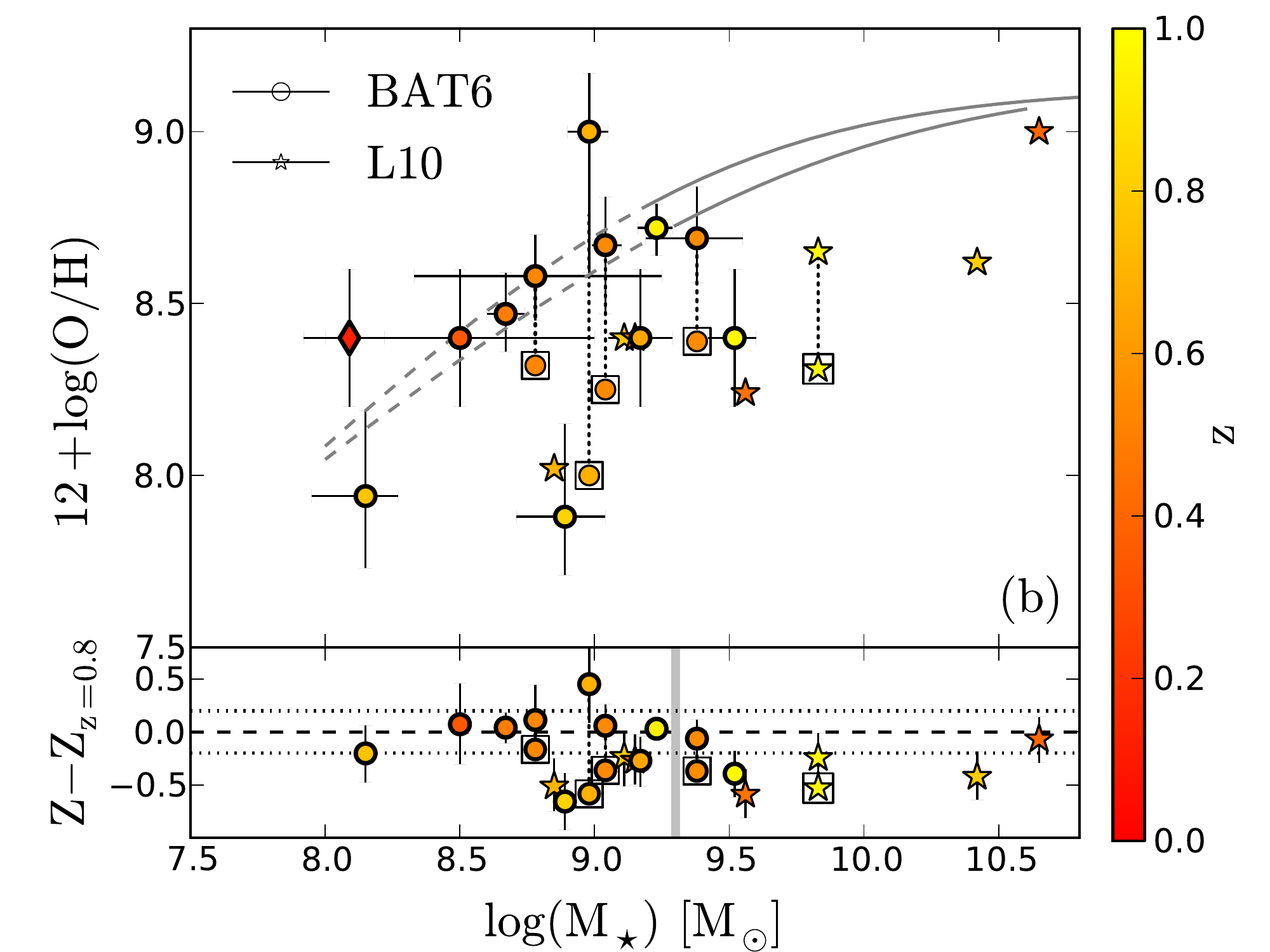}\\
 \end{tabular}
\caption{Comparison of the BAT6 sample hosts (circles) to the average mass-metallicity relations at different redshifts. {\it (a)} Metallicities are presented in \citet{Maiolino2008} calibration. Overplotted are the  models fitted to star-forming galaxy populations at different mean redshifts in the range of $z \sim 0.07 - 4$ \citep[M09; ][]{Mannucci2009}. As a comparison sample (stars) we plot the incomplete sample compiled by \citet{Mannucci2011} over $0.3<z<1$. {\it (b)} Metallicities are presented in \citet{Kobulnicky2004} calibration. Both upper and lower branch solution are plotted in cases where one solution cannot be obtained - in these cases the two values are connected with a dashed line and lower branch solution is plotted within a square for clarity. For comparison we also include the incomplete sample of LGRB hosts from \citet{Levesque2010} (stars) $0.3<z<1$. Lines represent fitted relations for galaxies at $z = 0.3$ and 0.8 \citep{Zahid2013a}. The extrapolation towards low stellar masses is indicated by dashed lines. Lower panels show the difference between the LGRB metallicities ($0.3<z<1$) and the median relations at redshift (a) 0.7 and (b) 0.8, respectively. Vertical grey lines in the lower panels mark the mass below which the two relations have been extrapolated. Errors of the comparison samples are not plotted in upper panels for clarity, but are taken into account when calculating the difference from median relations (both errors in mass and metallicity are accounted for). Dotted horizontal lines in lower panels show intrinsic dispersion of the median relations - we assume a typical value of $\pm 0.2$ dex.}
 \label{plot4a}
\end{figure*}

\begin{figure*}[t]
\centering
\begin{tabular}{cc}
\includegraphics[scale=0.45]{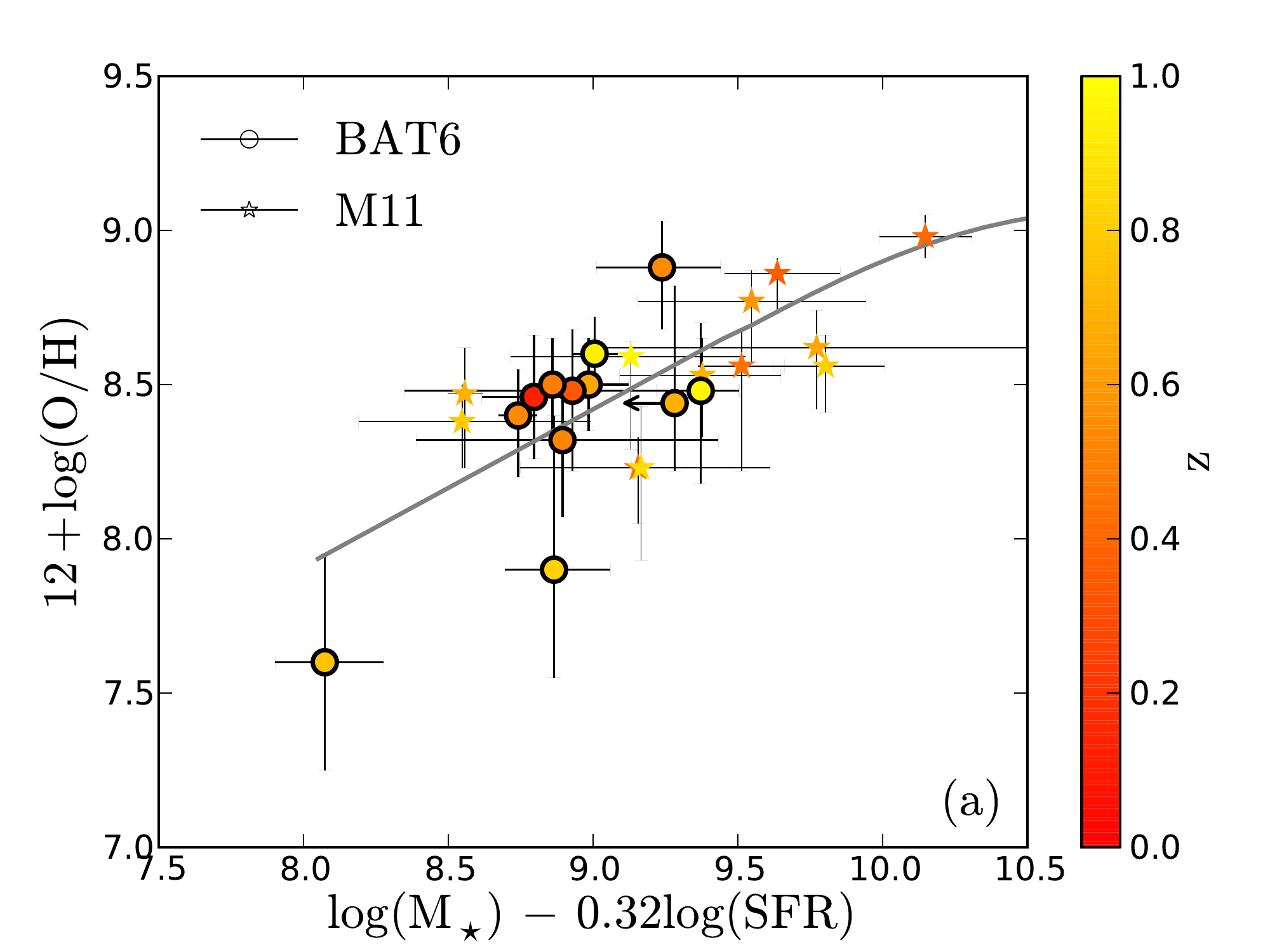}&
\includegraphics[scale=0.34]{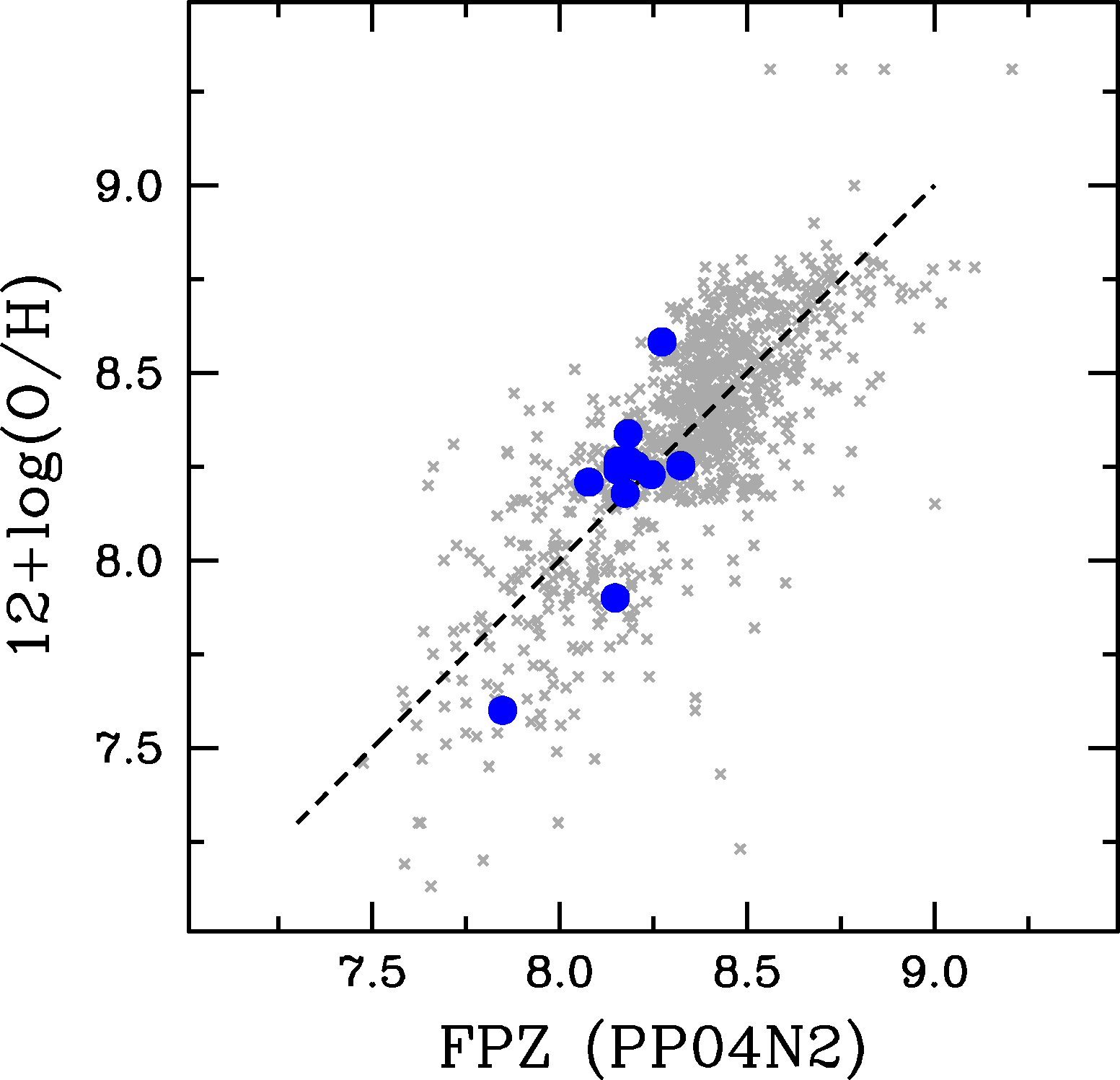}\\
 \end{tabular}
\caption{{\it (a)} Fundamental metallicity relation \citep{Mannucci2010,Mannucci2011}. Our sample (circles) is compared to the incomplete sample of \citet{Mannucci2011} (stars). {\it (b)} Fundamental metallicity plane (FPZ) for low-mass galaxies \citep{Hunt2016}. BAT6 sample (blue) is compared to different species of low-mass galaxies (grey). The plotted relation is done with \NII/H$\alpha$ metallicity calibration \citep[N2;][]{Pettini2004}, and our data have been transformed to this calibration following \citet{Kewley2008}.
}
 \label{plot4b}
\end{figure*}

\subsubsection{Comparison with starbursts and BCD galaxies}
\label{starburst}

We further compare the SFR and sSFR vs stellar mass trend with the $0.3<z<1$ star-forming galaxies studied by \cite{Atek2014} and the low-mass star forming galaxies and BCDs studied in \cite{Rodriguez2015} (see Section \ref{othersample}). GRB host galaxies have on average higher stellar masses than BCDs. We cannot compare the stellar masses with low-mass star forming galaxies of \cite{Rodriguez2015} as they were originally selected to have stellar masses less than 10$^8$\,M$_{\sun}$, i.e. the mass region that is not covered by our BAT6 sample. In the common covered range of stellar masses, SFR and sSFR show a large but similar spread. The selection of the \cite{Atek2014} sample was based on emission line detections, therefore its SFR limit ($\log ~{\rm SFR} {\rm [M}_{\odot} {\rm yr}^{-1}{\rm ]}\sim-1$) should be taken into account when comparing with our LGRB host galaxy sample. Within the SFR limits of the \cite{Atek2014} surveys (therefore excluding the host galaxies of GRB\,060614 and GRB\,061021), GRB host galaxies occupy a smaller stellar mass range and have similar SFR. 

Since the \citet{Atek2014} survey specifically selected galaxies with high specific star-formation rates, we became interested in the percentage of starbursts in the LGRB hosts, \citet{Atek2014} and field (VVDS) galaxy samples in the $0.5 < z < 1.0$ redshift range - the average redshift of each of the three populations in this range is $z \sim 0.7$. \citet{Whitaker2012} studied a sample of $0 < z < 2.5$ star-forming galaxies and found that the SFR-M$_{\star}$ relation of their sample had a scatter of $\pm 0.34$ around the median relation and that the scatter was independent of stellar mass and redshift. The median relation (equation (1) in \citealt{Whitaker2012}) at $z = 0.7$ is plotted in Figures \ref{fig2} and \ref{fig3}. Following their result we calculated how many galaxies of a given population have specific star-formation rate above the star-formation sequence (e.g. above the dotted line indicating the +0.34 dex scatter; see Figures \ref{fig2}b and \ref{fig3}b). We found that 27 (-9,+15)$\%$, 27$\%$ and 17$\%$ of galaxies are catagorized as starburst (according to our prescription) in the case of LGRB hosts, \citet{Atek2014} sample and the VVDS field sample, respectively\footnote{ \citet{Whitaker2014} extends the analysis to lower stellar masses. However, their analysis prevents them to study the scatter around the median relation, therefore we cannot use their findings for the present study. We note that, assuming the median SFR-M$_{\star}$ relation by \citealt{Whitaker2014} and a constant scatter of $\pm 0.34$, the fractions of starbursts do not change significantly.}. This result is in line with our expectations. Indeed, as the GRB formation probability scales, in some way, with the SFR, we should expect a larger incidence of starburst galaxies in GRB-selected samples with respect to those of field galaxies. It would be interesting to perform a similar analysis and to compare the SSFR cumulative distributions taking into account the completeness limits of the survey. Unfortunately we lack the statistics as our sample will be reduced to 6 objects only, therefore useless to obtain robust results.

\subsection{Metallicities}

\subsubsection{Mass-metallicity relation}
\label{MZrelation}

In Figure \ref{plot4a} we plot the MZ relation of the BAT6 sample both in the (a) M08 and (b) KK04 metallicity calibration. 

\begin{figure}[t]
\centering
\includegraphics[scale=0.5]{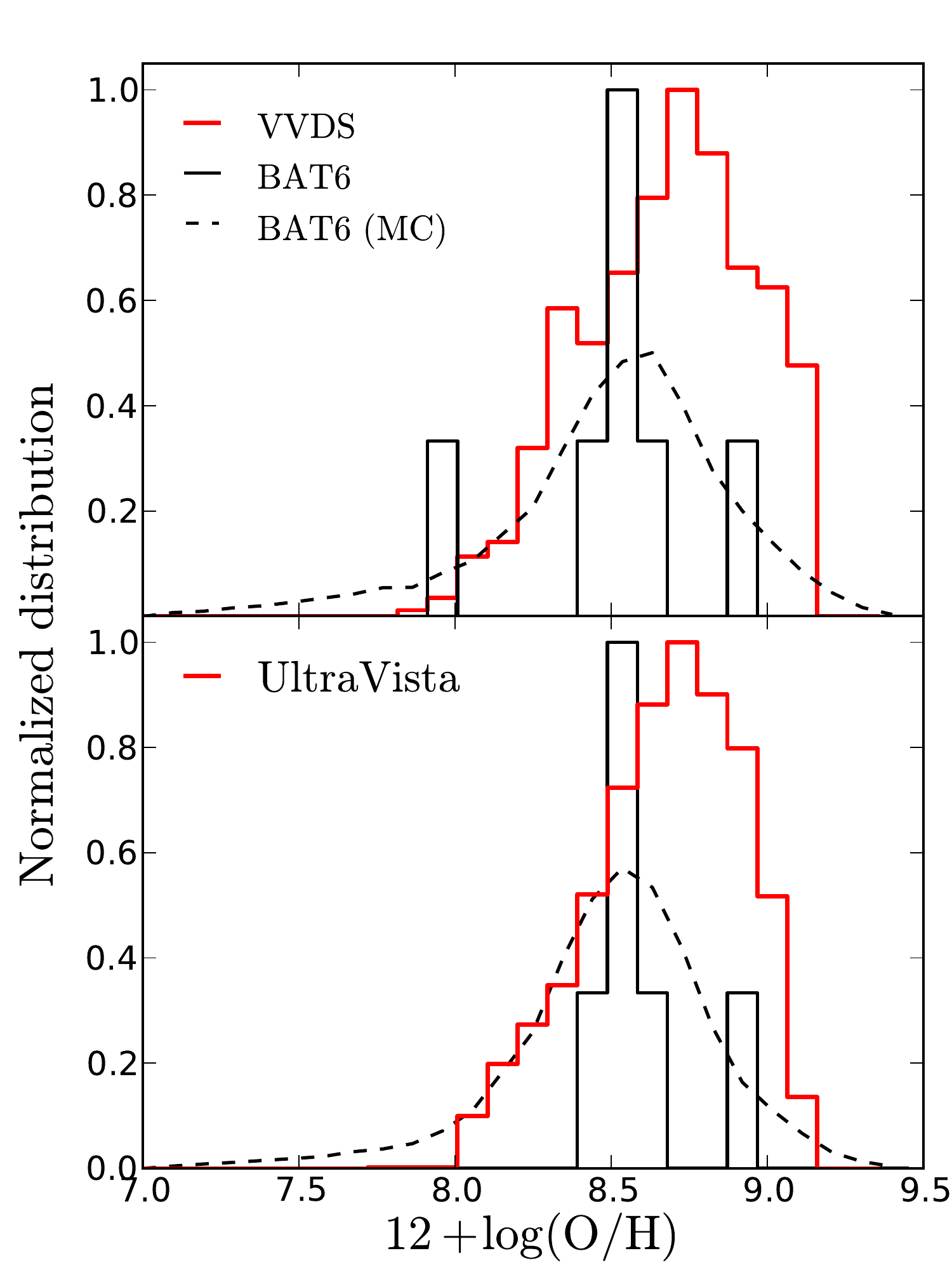}
\caption{Metallicity distribution of the BAT6 sample (solid black line) compared to the SFR-weighted distributions of VVDS and UltraVista samples of field galaxies (red lines). Dashed lines represent the average BAT6 distribution, obtained by taking into account the errors of the measured metallicities (via MC simulation). 
}
 \label{plot5}
\end{figure}  

We can compare our values with the MZ relation of field galaxies, taking into account the evolution of the relation with redshift
\citep[e.g.][]{Savaglio2005,Mannucci2009,Zahid2013a}. If we exclude the host of troublesome GRB\,060614, the redshift range of our sample is $0.3<z<1$ and the average redshift is $\sim 0.7$. Therefore we compare our sample to the median value of MZ relation at this redshift bin. The paucity of LGRB host galaxies at super-solar metallicity is evident. Accounting for errors, the fraction of hosts with metallicities above solar is found to be $16 (-8,+16) \%$. For comparison, \citet{Kruhler2015} retrieve the same result of $16 \pm 7 \%$ for their sample of $z < 1$ hosts. 
At sub-solar metallicities, our sample appears fairly consistent with the MZ relation within the dispersion and it does not show a systematic shift towards values below the relation found in some of the incomplete samples (see e.g. Figure \ref{plot4a}b and the sample of \citealt{Levesque2010}). For four hosts we could not break the degeneracy of the KK04-based metallicity. If the lower-branch solution is assumed as the correct one for the four cases, then our sample seems to follow the MZ relation up to $\log {\rm M}_{\star}{\rm [M}_{\odot}{\rm ]}\sim 8.7$, after which it starts to deviate towards lower metallicities. In the latter case it behaves in a similar way as the one from \citet{Levesque2010} over the same redshift range. Regardless of interpretation, we note that the comparison of our sample with the star-forming MZ relation is subjected to much uncertainty, as many of our hosts have masses below the limits of MZ relation obtained with galaxy surveys. We simply extrapolated the polynomials fitted to the relations at higher masses, however this may deviate from the real conditions. We also briefly mention that similar conclusions as for the MZ relation can be found when considering the relation between gas-phase metallicity and stellar-to-gas mass ratio, as parametrized by \citet{Zahid2014}. We show this in Figure \ref{plot12} in the Appendix.

It has been shown that star-forming galaxies (at least up to $z < 2.5$) follow a well defined relation between stellar mass, SFR and metallicity, known as the fundamental metallicity relation (FMR; \citealt{Mannucci2010}). We plot the FMR relation for GRB hosts in Figure \ref{plot4b}a. As found by \citet{Mannucci2011}, LGRB host galaxies follow the FMR within errors, i.e. are equally scattered around the relation even if with a quite large dispersion. Nevertheless, FMR is not well defined when approaching low stellar masses dominated by our sample. For example, \citet{Hunt2012} found that low-mass starburst galaxies deviate from the relation and that the mass-SFR-Z plane has to be recalibrated for such objects. We therefore also plot LGRB hosts in the recently calibrated relation \citep{Hunt2016}. As shown in Figure \ref{plot4b}b, LGRB hosts lie near the relation with a similar scatter as other low-mass galaxy samples.

\begin{figure*}[t]
\centering
\begin{tabular}{cc}
\includegraphics[scale=0.45]{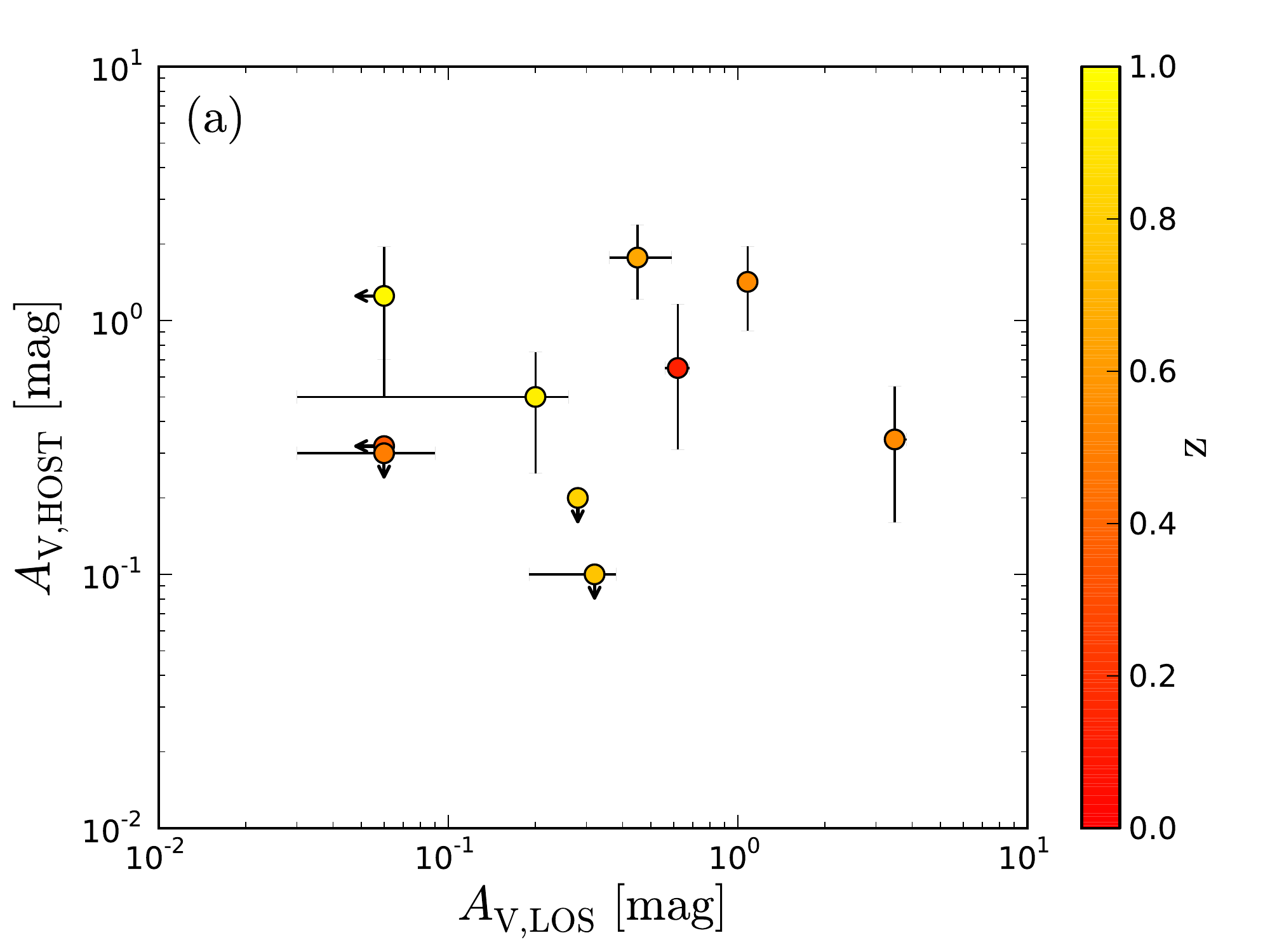}&
\includegraphics[scale=0.45]{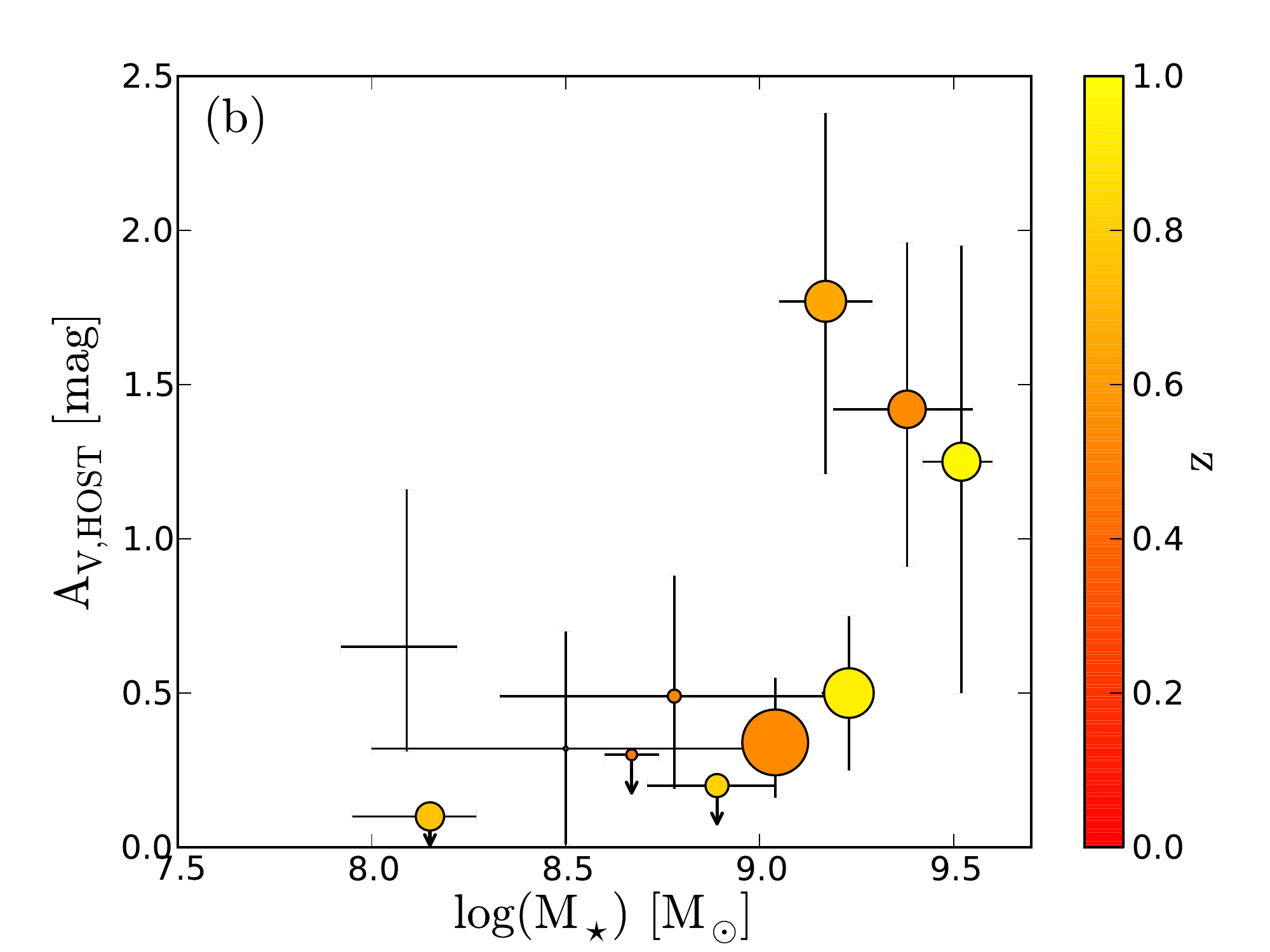}\\
 \end{tabular}
\caption{{\it (a)} Host-averaged extinction, measured from Balmer lines (e.g. Table \ref{tab_result}) compared to line-of-sight extinction \citep{Covino2013}. Color-coding corresponds to redshift. {\it (b)} Observed relation between stellar mass and host-averaged extinction. Circle sizes for the hosts are proportional to the values of their star formation rates.}
 \label{plot_ext}
\end{figure*} 

\begin{figure*}[t]
\centering
\includegraphics[scale=0.45]{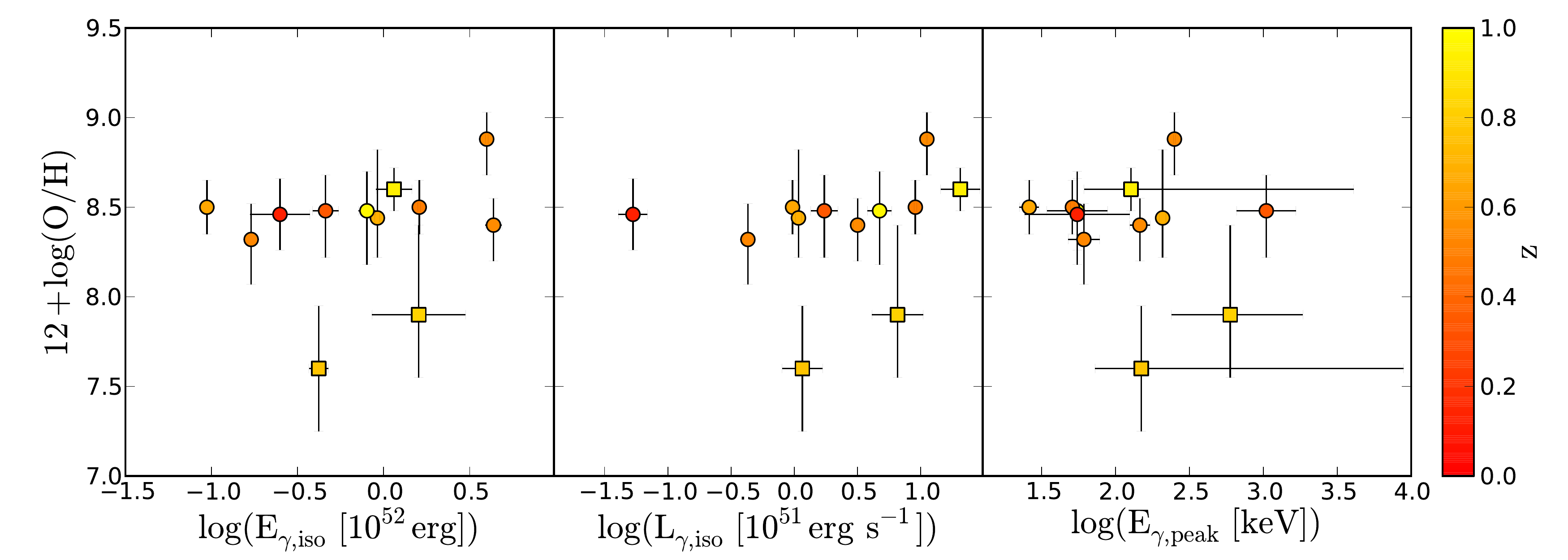}
\caption{Comparing high-energy properties of GRBs in the BAT sample and metallicities of their host galaxies. High energy properties of events plotted with circles are taken from \citet{Nava2012}. The three events plotted with squares have their peak energy (i.e. the peak in $\nu F_{\rm \nu}$ spectrum) outside the energy coverage of the BAT instrument: we estimated the peak by using the correlation between $E_{\rm \gamma,peak}$ and spectral index $\Gamma$, found by \citet{Sakamoto2009}, and then computed $E_{\rm \gamma,iso}$ and $L_{\rm \gamma,iso}$ in the extrapolated 1-10000 keV range \citep[e.g.][]{Pescalli2015}.}
 \label{plot7}
\end{figure*} 

\subsubsection{Metallicity distribution}
\label{Zdist}

With the MZ relation one can examine whether the LGRB host population lies in the same plane as star-forming galaxies, but does not give insight on the frequency with which LGRBs occur as a function of metallicity with respect to the star-forming population. Therefore we also compare metallicity distribution of BAT6 hosts to VVDS and UltraVista (see paper I) field galaxy samples (Figure \ref{plot5}). For both surveys we calculated the metallicity using the FMR relation. Since it is based on the stellar mass and SFR of the galaxies, we apply a cut to our GRB host galaxy sample following the stellar mass and SFR limits of both surveys to make the comparison. In the case of VVDS, we select events with $ i{\rm (AB)} < 24.75$ and $\log {\rm SFR} {\rm [M}_{\odot} {\rm yr}^{-1}{\rm ]} > 0.0$, while for the UltraVista comparison the selection encompass events with $K{\rm(AB) < 24.0}$ and $\log {\rm SFR} {\rm [M}_{\odot} {\rm yr}^{-1}{\rm ]}> 0.4$. The resulting BAT6 samples are therefore cut to a rather low number of 6 and 7 events, respectively. We note that the BAT6 hosts without measured metallicities (hosts of GRBs\,050525A and 080319B) and the host of peculiar GRB\,060614 are automatically excluded from the comparison samples, thus not affecting the conclusions. The comparison samples are furthermore limited to $z > 0.3$, both due to incompleteness of the two surveys at $z \lesssim 0.3$ as well as the lack of GRB hosts in BAT6 sample in that redshift range. 

The samples compared in Figure \ref{plot5} have similar average redshifts: both VVDS and UltraVista samples have $<z> = 0.76$, while BAT6 samples have $<z> = 0.74,0.72$ in the top and bottom plot. As we are dealing with a small sample of LGRB hosts, we built a median distribution (dashed line) by taking errors into account and perform a MC simulation. Our comparisons to VVDS and UltraVista surveys, taking into account the completeness in brightness, SFR and M$_{*}$ of the samples, indicate that the metallicities of LGRB hosts and star-forming galaxies have similar distributions up to $12 + \log \left( \frac{\rm O}{\rm H}\right) \sim 8.4-8.5$, after which the already discussed paucity of high-metallicity hosts is observed. This cutoff value, obtained by direct comparison, is similar to the one found in an indirect way in complete-sample studies by \citet{Vergani2015} and \citet{Perley2015b}. Finally, we note that there are two hosts with very low metallicities in the BAT6 sample with values which are not well constrained, thus limiting their weight in the analysis. Also, the small number of events used in the comparison prevents us from making quantitative, and therefore stronger, statistical conclusions.

\subsection{Dust}
\label{secdust}

Lastly we examine the dust properties of our host sample. We start by checking the relation between host-averaged extinction ($A_{\rm V,HOST}$), measured from Balmer decrement (i.e.\,Table \ref{tab_result}), and extinction in the GRB line-of-sight ($A_{\rm V,LOS}$), measured from the SED analysis \citep{Covino2013}. Out of 14 events in the BAT6 sample, 10 cases have measurements (or estimated upper limits) of both quantities. Figure \ref{plot_ext}a reveals that the two quantities of our sample of hosts are not correlated. \citet{Perley2013} looked at the same relation using a sample of GRBs extending to higher redshifts and higher line-of-sight extinctions, i.e. their work focused on the class of dark bursts \citep[e.g.][]{Jakobsson2004}. They showed that, approximately, the more extinct afterglows indeed tend to originate in dustier hosts. However, the relation is subject to considerable deviations of individual bursts from the $A_{\rm V,HOST} = A_{\rm V,LOS}$ correspondence and, especially for low $A_{\rm V,LOS}$, to large dispersion. Furthermore, at redshifts $z < 1$ \citet{Perley2013} do not find any host with $A_{\rm V,HOST}$ larger from the values in our sample. The lack of correlation for the $z < 1$ sample is thus consistent with previous studies.

It has been established that extinction in star-forming galaxies in general increases with stellar mass \citep[e.g.][]{Zahid2013}. We show in Figure \ref{plot_ext}b that the trend is also observed in our sample, though admittedly our analysis includes galaxies from a large redshift interval, in which the observed evolution of extinction with redshift could by itself introduce a bias into the relation. 

\subsection{High energy properties}

The BAT6 sample selection is based on the brightness of the prompt gamma-ray burst emission. To further verify the robustness of our results we thus check whether there is a correlation between the GRB energy output in $\gamma$-ray emission ($E_{\gamma}$) and its host metallicity. We look for a relation between metallicities and high energy properties, namely isotropic equivalent $\gamma$-ray energy $E_{\rm \gamma,iso}$, isotropic peak luminosity $L_{\rm \gamma,iso}$ and peak energy $E_{\rm \gamma,peak}$, for our BAT6 sample. We find no evidence for a correlation of these properties with metallicity. Similar conclusions have been also found by \citet{Levesque2010b}. With the present evidence we can therefore assume that our results are not affected by our sample selection criteria.

\section{Conclusions}
\label{conclude}

We have presented a spectroscopic study of a sample of 14 $z < 1$ LGRB host galaxies drawn from the {\it Swift}/BAT6 complete sample of bright LGRBs. Our work compares derived host galaxy properties (SFR, metallicity, stellar masses) to those of the general star-forming galaxy population, as well as investigating relations between those properties. 

We investigate the role of metallicity in the efficiency of LGRB production. Early studies (see Introduction) on the subject based on incomplete LGRB samples reported strong preference toward low metallicity values. Lately, however, various studies have indirectly pointed out that this view is only partially correct \citep{Vergani2015,Kruhler2015,Perley2015b}. Our results show that at $0.3<z<1$ LGRBs preferentially select galaxies of sub-solar metallicities ($12 + \log \left( \frac{\rm O}{\rm H}\right) \sim 8.4-8.5$) and therefore of low stellar masses. While the paucity of the super-solar metallicity hosts is striking, at sub-solar metallicities we find no evidence for a shift towards lower values on the MZ relation based on star-forming galaxies at similar redshift.

The preference for LGRBs to explode in sub-solar metallicity galaxies is likely also the explanation of the observational evidence that LGRB hosts at $z<1$ have on average lower star-formation rates than if they were direct star-formation tracers. Nevertheless, within the population of low-metallicity, low-mass and low-SFR galaxies they seem to be preferentially selecting galaxies with high SFR, as shown by an increased fraction of starbursts (i.e. high specific SFR galaxies) among the LGRB host galaxies with respect to those of the field star-forming galaxy population. Unfortunately our sample is too small (and galaxy surveys not deep enough) to obtain a robust result and to investigate more if the starburst fraction of host galaxy is the one expected under the hypothesis that GRBs are connected to SFR, after taking the high metallicity aversion into account.

The preference for LGRBs to avoid high metallicity galaxies can be related to the condition necessary for the progenitor star to produce a LGRB. Single star progenitor models favor low metallicity, but some of them require very low metallicity cuts \citep{Hirschi2005,Yoon2006} and cannot explain hosts with observed near-solar (or higher) metallicity. All resolved host galaxy observations have shown that LGRB host galaxies have almost negligible metallicity gradients \citep[e.g.][]{Christensen2008,Levesque2011}. Assuming that this holds for all hosts, the discrepancy between the expected low metallicity cut and observed near solar metallicity therefore cannot be explained by the difference between the metallicity at the explosion site and the measured host-averaged metallicity. Furthermore, \citet{Modjaz2008} found that broad-line core-collapse SNe accompanying LGRBs are found in less metal-rich environment compared to those without detected GRBs, with a metallicity threshold similar to the one found in this study. This result shows that the two types of transients preferentially occur in different conditions and suggests different progenitor properties. There is more and more evidence that binary stars represent a significant fraction of core-collapse SN progenitors. Even if at a lesser extent, metallicity can influence also the evolution of binary stars (Belczynski, private communication). It will be interesting in the future to compare our results with some quantitative predictions of the metallicities of binary stars as LGRB progenitors.

We emphasize that despite the vast and rich existing literature on star-forming galaxies, it was not easy to find comparison field galaxy samples whose completeness limits would be suitable for comparison to LGRB hosts. Even at low and intermediate redshifts LGRB host population can thus be complementary to surveys studying the low-mass, faint galaxy population, in particular when extending mass-metallicity (or FMR) relation to low stellar masses ($\sim 10^{8}$ M$_{\odot}$). LGRBs preferentially select metal-poor galaxies. It has been suggested that M$_{\star}$ and SFR are the main parameters driving the FMR, not metallicity \citep{Hunt2012}. Therefore LGRBs select low-mass galaxies much more effectively than magnitude-limited surveys.

Albeit the sample of galaxies used in this study is small, we emphasize the importance of using complete samples to understand the properties of the LGRB host population. In the future we will move our analysis towards higher redshifts with the aim of obtaining a deeper insight into the condition affecting the rate of LGRBs to confirm if, as suggested by recent studies \citep{Greiner2015,Perley2015b}, LGRBs become direct SFR tracers as we move back through cosmic time. 

\begin{acknowledgements}
We thank the referee for the helpful comments for improving the paper. SDV thanks C. Belczynski, C.Georgy, J. Groh, O. Le Fèvre, L. Kewley and L. Tasca for fruitful discussions. JJ and SC acknowledge financial contribution from the grant PRIN MIUR 2012 201278X4FL 002 The Intergalactic Medium as a probe of the growth of cosmic structures. SDV and ELF acknowledge the UnivEarthS Labex programme at Sorbonne Paris Cité (ANR-10-LABX-0023 and ANR-11-IDEX-0005-02). AFS acknowledges support from grants AYA2013-48623-C2-2 from the Spanish Ministerio de Economia y Competitividad, and PrometeoII 2014/060 from the Generalitat Valenciana. This research uses data from the VIMOS VLT Deep Survey, obtained from the VVDS database operated by Cesam, Laboratoire d'Astrophysique de Marseille, France. This work is partly based on observations made with the Gran Telescopio Canarias (GTC), installed in the Spanish Observatorio del Roque de los Muchachos of the Instituto de Astrofísica de Canarias in the island of La Palma.
\end{acknowledgements}


\bibliographystyle{aa}
\bibliography{hosts_bat6_bib}
\appendix
\renewcommand{\thesection}{A.\arabic{section}}
\section{Appendix A}

\begin{figure*}[t]
\centering
\includegraphics[scale=0.6]{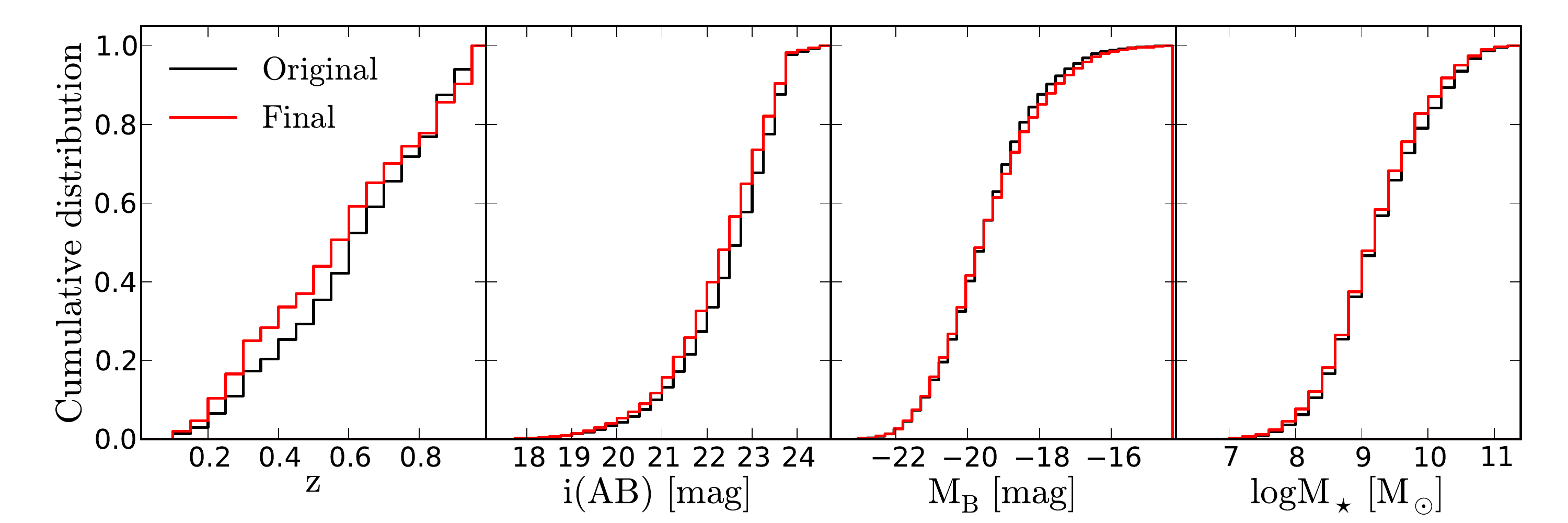}
\caption{Cumulative plots of properties of the comparison galaxy sample from the VVDS survey illustrating that the original sample (6366 galaxies, black lines) and the final sample (3551 galaxies, red lines) that was used in the analysis, do not differ in their properties and that no bias was introduced throughout the selection process.}
\label{plot11}
\end{figure*}

\begin{figure*}[t]
\centering
\includegraphics[scale=0.45]{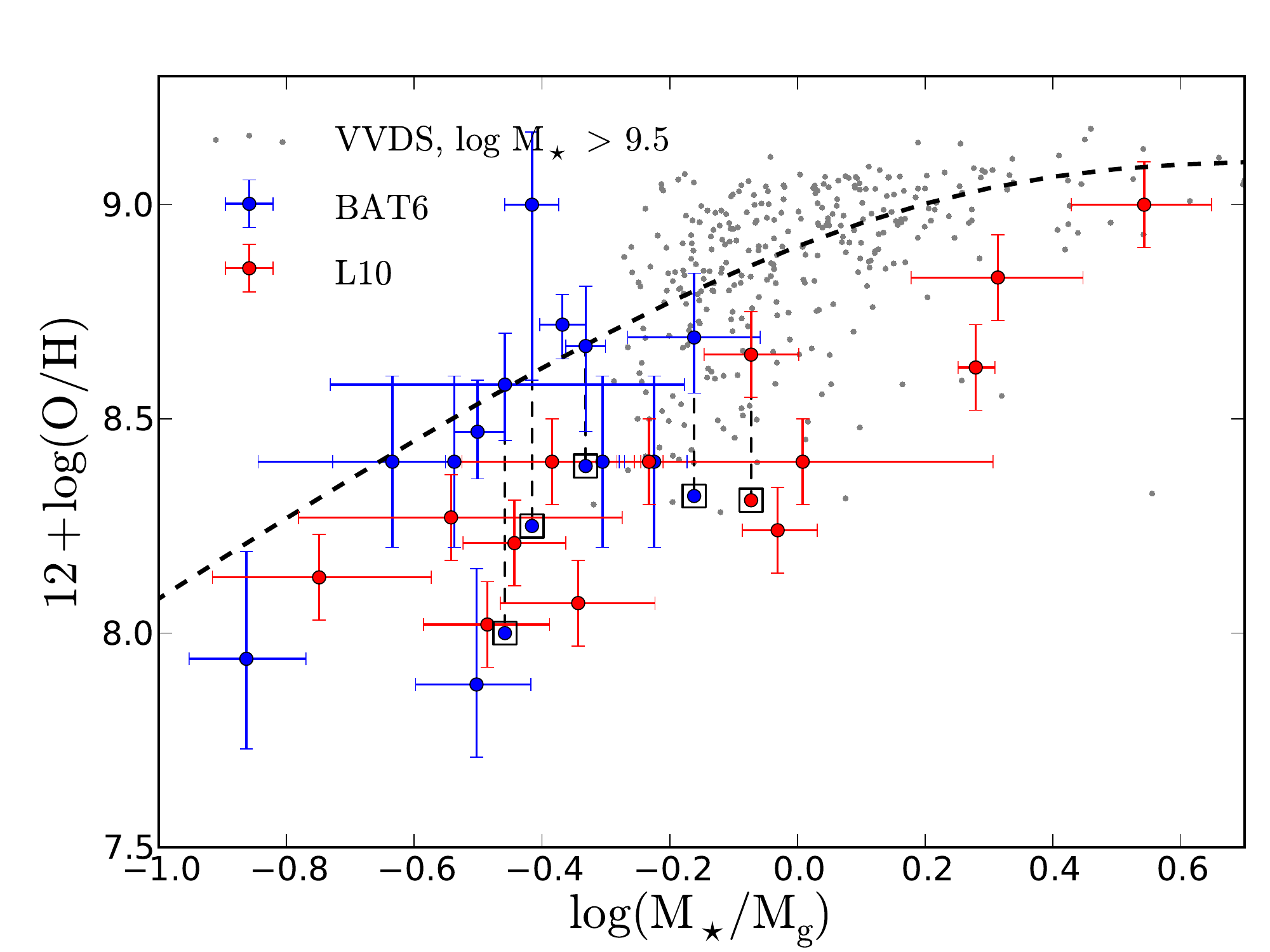}
\caption{Relation between metallicity and stellar-to-gas mass ratio. Metallicity is given in the calibration of \citet{Kobulnicky2004} (KK04). The dashed line is the relation found by \citet{Zahid2014} for star-forming galaxies. Gas (hydrogen) masses are computed using Eq. 34 in \citet{Zahid2014}.  Errors on the $x$-axis only take into account the errors on measured M$_{\star}$: the true errors are larger due to error in metallicity calibration ($\sim 0.15$ dex) and the scatter in the relation itself ($\sim 0.07$ dex). Blue data are BAT6 sample, red represent \citet{Levesque2010} sample and grey dots values computed from VVDS sample (only $\log$ M$_{\star} > 9.5$, which allows an assumption that all correct KK04 values are of the upper branch). It is evident that the biased sample used by \citet{Levesque2010} lies below the median relation of star forming galaxies. While a few of our own LGRB hosts seem to be outliers, most of our hosts are consistent within errors with the relation.}
\label{plot12}
\end{figure*}

\begin{landscape}
\begin{table}
\renewcommand{\arraystretch}{1.3}
\begin{center}
\small
\begin{tabular}{lcccccccccc}
\hline
\hline
GRB & \multicolumn{2}{c}{\OII$^{(a)}$}      & \NeIII & H$\delta$ & H$\gamma$ & H$\beta$ & \multicolumn{2}{c}{\OIII} & H$\alpha$ & \NII\\
\cmidrule(lr{.75em}){2-3}
\cmidrule(lr{.75em}){8-9}
       & $\lambda3726$ & $\lambda3729$ &          &                  &                     &                & $\lambda4959$  & $\lambda5007$         &                  &
$\lambda6584$ \\
\hline
050416A & 4.3 $\pm$ 0.5 & 4.1 $\pm$ 0.5 & $<$ 1.3 & $<$ 1.0 & 1.1 $\pm$ 0.3 & 1.9 $\pm$ 0.4 & 1.8 $\pm$ 0.5 & 6.3 $\pm$ 0.5 & 11.8 $\pm$ 1.4 & 0.8 $\pm$ 0.3\\
050525A$^{(i)}$ & $<$ 1.7 & $<$ 1.7 & $<$ 2.0 & $<$ 1.6 & $<$ 2.0 & $(s)$ & 1.2 $\pm$ 0.4 & 2.7 $\pm$ 0.4 & 2.2$^{(h)}$ & $<$ 2.5\\
060614 & \multicolumn{2}{c}{2.8 $\pm$ 0.5} & $<$ 0.7 & $<$ 1.2 & $<$ 0.8 & 0.7 $\pm$ 0.2 & $(s)$ & 2.8 $\pm$ 0.7 & 2.5 $\pm$ 0.5 & $<$ 1.0 \\
060912A & 7.3 $\pm$ 0.4 & 10.8 $\pm$ 0.5 & 1.8 $\pm 0.5$ & 0.7 $\pm$ 0.3 & 2.5 $\pm$ 0.3 & 5.3 $\pm$ 0.5 & 4.0 $\pm$ 0.5$^{(t)}$ & 11.2 $\pm$ 1.2 & 17.4 $\pm$ 2.0 & 3.6 $\pm$ 2.5\\
061021 & 1.0 $\pm$ 0.2 & 1.0 $\pm$ 0.2 & $<$ 0.6 & $<$ 0.8 & $<$ 2.1 & 0.5 $\pm$ 0.1 & 0.6 $\pm$ 0.1 & 1.7 $\pm$ 0.3 & 1.6 $\pm$ 0.3 & $<$ 0.5\\
071112C$^{\dagger}$ & \multicolumn{2}{c}{1.9 $\pm$ 0.4} & 0.6 $\pm$ 0.2$^{(s)}$ & 0.8 $\pm$ 0.4 & 1.2 $\pm$ 0.4 & $<$ 2.0 & 2.5 $\pm$ 0.5 & 2.4 $\pm$ 0.5$^{(t)}$ & & \\
080430  & \multicolumn{2}{c}{3.5 $\pm$ 0.8} & 2.3 $\pm$ 0.5 & 3.0 $\pm$ 0.7 & $(t)$ & 4.7 $\pm$ 1.8 & 5.5 $\pm$ 1.2 & $(s)$ & & \\
080916A &\multicolumn{2}{c}{2.9 $\pm$ 0.4} & $(s)$ & $<$ 0.4 & $<$ 0.5 & 1.8 $\pm$ 0.7$^{(t)}$ & $(t)$ & 3.6 $\pm$ 0.6 & & \\
081007  & 1.3 $\pm$ 0.2 & 1.6 $\pm$ 0.2 & $<$ 1.4 & $<$ 1.4 & 0.8 $\pm$ 0.2 & 1.4 $\pm$ 0.2 & 1.7 $\pm$ 0.2 & 5.9 $\pm$ 0.3 & 5.0 $\pm$ 0.3 & $(d)$\\
090424 & \multicolumn{2}{c}{13.8 $\pm$ 2.3}  & $(s)$ & $(e)$& 1.7 $\pm$ 0.4 & 4.3 $\pm$ 0.6 & 1.3 $\pm$ 0.5$^{(t)}$ & 3.5 $\pm$ 0.4 & 19.9 $\pm$ 3.3 & $(s)$\\
091018$^{(b,\dagger)}$ & 1.7 $\pm$ 0.2 & 2.9 $\pm$ 0.4 & $(t)$ & $<$ 0.8 & $<$ 0.8 & 1.3 $\pm$ 0.3$^{(t)}$  & 1.7 $\pm$ 0.3 & 4.6 $\pm$ 1.0 & 5.6 $\pm$ 1.6$^{(s)}$ & $<$ 1.5\\
091127$^{\dagger}$ & 4.2 $\pm$ 0.7 & 5.4 $\pm$ 0.9 & $<$ 1.4 & 0.7 $\pm$ 0.3 & 0.7 $\pm$ 0.2 & 2.3 $\pm$ 0.2 & 3.0 $\pm$ 0.2 & 7.8 $\pm$ 0.3 & 6.0 $\pm$ 0.3 & 0.5 $\pm$ 0.2\\
100621A & 38.4 $\pm$ 5.4 & 49.0 $\pm$ 6.0 &15.1 $\pm$ 3.2 & 11.4 $\pm$ 1.3 & 17.6 $\pm$ 1.3 & 38.0 $\pm$ 2.1 & 35.6 $\pm$ 1.1 & 130.4 $\pm$ 6.3 & 130.8 $\pm$ 11.5 & 9.7 $\pm$ 5.1$^{(s)}$ \\
\hline
\hline
\end{tabular}
\end{center}
\caption{Measured line fluxes ($10^{-17}$ erg cm$^{-2}$ s$^{-1}$), corrected for Galactic extinction and stellar Balmer absorption. Upper limits are estimated at 3$\sigma$ level. No values are reported in cases where a line lies outside the covered spectral range or if it falls in the region of strong telluric absorption and is not detected. \newline(a) If only one line is reported, the OII doublet is not resolved. \newline (b) H$\beta$ lies in a region a strong telluric absorption, but it is clearly detected. Telluric absorption was modeled with molecfit \citep{Smette2015,Kausch2015} using the spectrum of the afterglow (has sufficiently high S/N) and applied to the spectrum. The corrected spectrum was used to measure H$\beta$. \newline (t) Likely affected by (or, if no number, missing due to) significant telluric absorption. \newline(s) Likely contaminated (or, if no number, completely dominated) by sky emission line residuals. \newline(d) \NII lies on the extremely noisy edge of the VIS spectral arm. \newline(e) H$\delta$ is actually observed in absorption. Balmer correction for this host is substantial. \newline (h) The line is marginally detected but lies in a very noisy region. We measure its flux by fixing the centre and width of the Gaussian in the fit. The width was assumed to be the same as the one measured from the \OIII$\lambda5007$ line. \newline (i) We note that Kr\"uhler et al. (2015) did not apply any additional correction to flux calibration for this host, given that the continuum is not detected. Our values of emission line fluxes are consistent with theirs before we apply the correction (Section \ref{flux_cal}).\newline $\dagger$ Spectrum dominated by afterglow.}
\label{line_flux}
\end{table}
\end{landscape}

\end{document}